\documentclass{aa}  

\usepackage{graphicx}
\usepackage{ulem}
\usepackage{txfonts}
\usepackage[breaklinks,colorlinks,citecolor=blue]{hyperref}

\usepackage{lmodern}

\bibpunct{(}{)}{;}{a}{}{,} 

\begin{document} 

\newif\iffigs

\figstrue

\newcommand{\itii}[1]{{#1}}
\newcommand{\franta}[1]{\textbf{\color{green} #1}}
\newcommand{\itiitext}[1]{{#1}}

\newcommand{\eq}[1]{eq. (\ref{#1})}
\newcommand{\eqp}[1]{(eq. \ref{#1})}
\newcommand{\eqb}[2]{eq. (\ref{#1}) and eq. (\ref{#2})}
\newcommand{\eqc}[3]{eq. (\ref{#1}), eq. (\ref{#2}) and eq. (\ref{#3})}
\newcommand{\refs}[1]{Sect. \ref{#1}}
\newcommand{\reff}[1]{Fig. \ref{#1}}
\newcommand{\reft}[1]{Table \ref{#1}}

\newcommand{\datum} [1] { \noindent \\#1: \\}
\newcommand{\pol}[1]{\vspace{2mm} \noindent \\ \textbf{#1} \\}
\newcommand{\code}[1]{\texttt{#1}}
\newcommand{\figpan}[1]{{\sc {#1}}}

\newcommand{\sfe}{\mathrm{SFE}}
\newcommand{\nbdvi}{\textsc{nbody6} }
\newcommand{\nbdvid}{\textsc{nbody6}}
\newcommand{\mum}{$\; \mu \mathrm{m} \;$}
\newcommand{\rop}{$\rho$ Oph }
\newcommand{\HT}{$\mathrm{H}_2$}
\newcommand{\Halpha}{$\mathrm{H}\alpha \;$}
\newcommand{\HI}{H {\sc i} }
\newcommand{\HII}{H {\sc ii} }
\renewcommand{\deg}{$^\circ$}

\newcommand{\dd}{\mathrm{d}}
\newcommand{\acosh}{\mathrm{acosh}}
\newcommand{\sign}{\mathrm{sign}}
\newcommand{\cex}{\mathbf{e}_{x}}
\newcommand{\cey}{\mathbf{e}_{y}}
\newcommand{\cez}{\mathbf{e}_{z}}
\newcommand{\cer}{\mathbf{e}_{r}}
\newcommand{\ceR}{\mathbf{e}_{R}}

\newcommand{\llg}[1]{\log_{10}#1}
\newcommand{\pder}[2]{\frac{\partial #1}{\partial #2}}
\newcommand{\pderrow}[2]{\partial #1/\partial #2}
\newcommand{\nder}[2]{\frac{\dd #1}{\dd #2}}
\newcommand{\nderrow}[2]{{\dd #1}/{\dd #2}}

\newcommand{\Cmiii}{\, \mathrm{cm}^{-3}}
\newcommand{\Gcmii}{\, \mathrm{g.cm}^{-2}}
\newcommand{\Gcmiii}{\, \mathrm{g.cm}^{-3}}
\newcommand{\Kms}{\, \mathrm{km} \, \, \mathrm{s}^{-1}}
\newcommand{\Si}{\, \mathrm{s}^{-1}}
\newcommand{\Esi}{\, \mathrm{erg} \, \, \mathrm{s}^{-1}}
\newcommand{\Ee}{\, \mathrm{erg}}
\newcommand{\Yr}{\, \mathrm{yr}}
\newcommand{\Myr}{\, \mathrm{Myr}}
\newcommand{\Gyr}{\, \mathrm{Gyr}}
\newcommand{\Msun}{\, \mathrm{M}_{\odot}}
\newcommand{\Rsun}{\, \mathrm{R}_{\odot}}
\newcommand{\Pc}{\, \mathrm{pc}}
\newcommand{\Kpc}{\, \mathrm{kpc}}
\newcommand{\Sd}{\Msun \, \Pc^{-2}}
\newcommand{\Ev}{\, \mathrm{eV}}
\newcommand{\Kk}{\, \mathrm{K}}
\newcommand{\Au}{\, \mathrm{AU}}
\newcommand{\Mas}{\, \mu \mathrm{as}}

   \title{Tidal tails of open star clusters as probes to early gas expulsion I: A semi-analytic model}


   \author{Franti\v{s}ek Dinnbier
          \inst{1},
          Pavel Kroupa \inst{2,3}
          }

   \institute{I.Physikalisches Institut, Universit\"{a}t zu K\"{o}ln, Z\"{u}lpicher Strasse 77, D-50937 K\"{o}ln, Germany \\
             \email{dinnbier@ph1.uni-koeln.de}
         \and
             Helmholtz-Institut f\"{u}r Strahlen- und Kernphysik, University of Bonn, Nussallee 14-16, 53115 Bonn, Germany \\
             \email{pavel@astro.uni-bonn.de}
         \and
             Charles University in Prague, Faculty of Mathematics and Physics, Astronomical Institute, V Hole\v{s}ovi\v{c}k\'{a}ch 2, 180 00 Praha 8, Czech Republic
             }

   \titlerunning{A semi analytic model for tidal tails}
   \authorrunning{F. Dinnbier \& P. Kroupa}

   \date{Received August 25, 2019; accepted June 17, 2020}

 
  \abstract
   {Star clusters form out of the densest parts of infrared dark clouds. 
    The emergence of massive stars expels the residual gas, which has not formed stars yet.
    Gas expulsion lowers the gravitational potential of the embedded cluster, unbinding many of the cluster stars. 
    These stars then move on their own trajectories in the external gravitational field of the Galaxy, forming a tidal tail.}
   {We investigate, for the first time, the formation and evolution of the tidal tail forming due to 
    expulsion of primordial gas. 
    We contrast the morphology and kinematics of this tail with 
    that of another tidal tail which forms by gradual dynamical evaporation of the star cluster.
    We intend to provide predictions which can differentiate the dynamical origin of possibly observed tidal tails around 
    dynamically evolved (age $\gtrsim 100 \Myr$) galactic star clusters by the Gaia mission. 
    Such observations might estimate the fraction of the initial cluster population which gets released in the gas expulsion event. 
    The severity of the initial gas expulsion is given by the star formation efficiency and the time-scale of gas expulsion for the cluster 
    when it was still embedded in its natal gas.
    A study with a more extended parameter space of the initial conditions is performed in the follow up paper.}
   {We provide a semi-analytical model for the tail evolution. 
    The model is compared against direct numerical simulations using \nbdvid.}
   {Tidal tails released during gas expulsion have different kinematic properties than the tails 
    gradually forming due to evaporation, which have been extensively studied. 
    The gas expulsion tidal tail shows non-monotonic expansion with time, where longer epochs of expansion are 
    interspersed with shorter epochs of contraction. 
    The tail thickness and velocity dispersions strongly, but not exactly periodically, vary with time. 
    The times of minima of tail thickness and velocity dispersions are given only by the properties of the galactic potential, 
    and not by the properties of the cluster. 
    The estimates provided by the (semi-)analytical model for the extent of the tail, the minima of tail thickness, and 
    velocity dispersions are in a very good agreement with the \nbdvi simulations.
    This implies that the semi-analytic model can be used for estimating the properties of the gas expulsion tidal tail 
    for a cluster of a given age and orbital parameters without the necessity of performing numerical simulations.}
   {}


   \maketitle


%

\section{Introduction}

\label{sIntro}


The problem of star cluster formation has been addressed by two classes of simulations: 
hydrodynamic simulations and N-body simulations, but it still possesses many unanswered questions partly 
due to the complex interplay between the involved physical processes, and partly due to the 
required spatial and temporal resolution of such simulations.

Hydrodynamic simulations usually start from a collapsing cloud, which forms stars, and the stars 
impact the cloud by their feedback, which in turn opposes the cloud self-gravity and tends to disperse it, 
suppressing further star formation. 
The final star formation efficiency (SFE) as well as the masses of individual stars 
are then given by the regulation of accretion by the multitude of feedback processes. 
Current state-of-the art simulations are able to model the formation of only small star clusters (stellar mass of $\approx 200 \Msun$; 
\citealt{Bate2014}, see also \citealt{Bate2012}) which self-consistently form stars with a realistic stellar initial mass function (IMF), 
which suggests that they contain all important ingredients for star formation and feedback. 
Simulations aiming at more massive clusters have to resort to coarser approximations, 
for example, they assume the shape of the IMF (e.g. \citealt{Dale2011,Dale2012,Grudic2018,Haid2019}), or 
terminate star formation at some point, remove instantly all gas, and continue with 
a purely stellar simulation \citep{Fujii2015b,Fujii2015a}, 
and/or take into account only some of the important feedback mechanisms completely neglecting 
the others (e.g. \citealt{Gavagnin2017,Wall2019}).  
Although these simulations provide valuable insights into the impact of the feedback on the gaseous 
configuration, their predictive power for the cluster forming mechanisms, which leave imprints for example in 
the SFE and the dynamics of the natal cluster, is at best, limited. 
Particularly, hydrodynamic simulations give very different prediction to the dispersal of molecular clouds and the SFE even for clouds of 
comparable masses (see for example \citealt{Dale2011,Dale2012} with \citealt{Gavagnin2017,Haid2019}). 

While hydrodynamic simulations focus on the gaseous phenomena and the stellar dynamics of the young 
cluster is usually only roughly approximated, N-body simulations adopt a complementary approach: 
they focus on detailed N-body dynamics at the expense of a rough approximation of gas. 
In N-body simulations, the gas distribution and its evolution is typically described by an analytic 
formula characterised by several parameters (e.g. the SFE, the time-scale of gas removal $\tau_{\rm M}$). 
Since the observational or theoretical constraints on some 
of these parameters are weak, they are taken as free parameters, and their 
plausible value is estimated by matching the evolved clusters with observed ones, 
effectively inverting the problem. 
In this way, N-body simulations can provide us with a clue not only to the initial conditions of the stellar component, 
but also to the gas component of natal star clusters.

Basic dynamical arguments distinguish between two different modes of gas expulsion, 
depending on the duration of the gas expulsion time-scale $\tau_{\rm M}$ relative to the half-mass crossing time 
$t_{\rm h}$ \citep{Hills1980,Mathieu1983,Arnold1989}, see also 
\citet[][chapt. 3.6]{Binney2008}. 
If the gas is expelled rapidly (impulsively, i.e. $\tau_{\rm M} \lesssim t_{\rm h}$), the cluster is generally more impacted than 
in the case of slow (adiabatic, i.e. $t_{\rm h} \ll \tau_{\rm M}$) gas expulsion, 
where the cluster gradually expands \citep[e.g.][]{Lada1984,Goodwin1997,Kroupa2001b,Baumgardt2007}.
The time-scale of gas expulsion influences, for example, the minimum value of the SFE which 
results in a gravitationally bound star cluster. 
While rapid gas expulsion necessitates the SFE to be at least $\approx 30$ \% for formation of a gravitational bound star cluster
(\citealt{Lada1984,Geyer2001,Kroupa2001b} see also the semi-analytic study of \citealt{Boily2003a,Boily2003b}), 
adiabatic gas expulsion enables cluster formation for substantially lower SFEs, 
typically $5-10$ \% \citep{Lada1984,Baumgardt2007}. 
The existence of evolved bound star clusters (e.g. the Pleiades), which likely contained O stars (these stars expel residual gas 
quickly due to their powerful feedback), sets a lower limit of the SFE in these systems to $\approx 30$\% \citep{Lada2003}. 

Interestingly, some results of N-body simulations seem to contradict the claims of very high SFEs
for more massive clusters (mass $\gtrsim 10^4 \Msun$) reported in some hydrodynamic simulations.
For example, the rapid expansion of star clusters from their initial subparsec radii \citep{Kuhn2014,Traficante2015} 
by a factor of at least $10$ \citep[][and references therein]{Pfalzner2009,Zwart2010} in $\approx10 \Myr$ is not likely to be explained 
by purely stellar dynamics, but it can be caused by rapid gas expulsion with relatively low SFEs of $\approx 30$\% \citep{Banerjee2017}. 
Also, the approximate virial state of some very young clusters (e.g. R~136; \citealt{Henault2012}) does not necessarily imply 
high SFEs because massive clusters revirialise rapidly after gas expulsion \citep{Banerjee2013}. 

Some models suggest a relatively low SFE and rapid gas expulsion even in such massive objects as were progenitors of globular clusters.
Low SFEs can explain the observed relation \citep{DeMarchi2007} between the
concentration of globular star clusters and the slope of their mass function \citep{Marks2008} as well as the 
unusual mass function of globular cluster Palomar~14 \citep{Zonoozi2011} without the necessity to venture to non-canonical stellar IMFs. 
The formation of globular clusters with low SFEs can also naturally explain how the globular cluster initial 
mass function of slope $\approx-2$ transformed to the currently observed mass function of globular clusters avoiding  
the assumption of exotic globular cluster initial mass functions \citep{Vesperini1998,Kroupa2002,Parmentier2005,Baumgardt2008}. 

Observationally, the SFE is directly determined only for the nearest star-forming regions,
which contain only lower mass star clusters (mass typically at most several hundreds solar masses).
These groups and clusters have the SFE mostly in the range $10-30$ \% \citep{Lada2003,Megeath2016}. 
The expansion of the majority (at least 75\%, but more probably 85-90\%) of young star clusters, 
which was recently measured due to Gaia \citep{Kuhn2019}, presents another piece of evidence for gas expulsion. 
In addition, the observed offset between the mass function of dense molecular cores and
the initial stellar mass function \citep{Alves2007,Nutter2007}
can be reconciled if cores retain $\approx 30$\% of their initial mass during their transformation to stars \citep{Goodwin2008}.
For more massive clusters (mass $\gtrsim 10^4 \Msun$), the SFE is observationally less constrained with the possibility that 
it increases as the clusters consume the majority of their star forming gas \citep{Longmore2014}.

A star cluster impacted by rapid gas expulsion with a relatively low SFE of $\approx 1/3$ evolves as follows. 
First, the stars continue moving with their pre-gas expulsion speeds, but now in a lowered gravitational potential, 
which results in their outward motions, and expansion of all Lagrangian radii. 
For this SFE, around $2/3$ of all stars escape the cluster. 
The rest of the stars revirialise, and the Lagrangian radii shrink, with the inner Lagrangian radii 
decreasing prior to the outer Lagrangian radii. 
The Lagrangian radii after revirialisation are larger than the initial Lagrangian radii, 
so revirialisation leaves the cluster rarefied relative to its initial state.  
After revirialisation, the cluster evolves as a purely N-body system with evaporation of the lower mass 
stars and occasional ejections of predominantly more massive stars \citep{Baumgardt2003,Fujii2011,Perets2012,Oh2015}. 

In this case, the cluster loses $\approx 2/3$ of its stars due to gas expulsion during the first Myr  
of its existence, with subsequent evaporation and ejections taking place afterwards for hundreds of Myr, 
the cluster losing roughly one star per half-mass crossing time. 
Since galactic star clusters orbit the Galaxy, the escapers are subjected to the Coriolis force and 
therefore form two tails: the one composed of stars released due to
gas expulsion (hereafter tail I), and the other composed of stars released due to evaporation (hereafter tail II) 
\footnote{These were referred to, respectively, as group I and group II by \citet{Kroupa2001b}.}
.


The purpose of this work (which we present in this and following paper, Dinnbier \& Kroupa accepted; hereafter paper II)
is to constrain the possible range of the SFE by studying the 
morphology, kinematics and stellar populations of tidal tails released due to gas expulsion. 
We take the SFE and the gas expulsion time-scale $\tau_{\rm M}$ as free parameters, and study the properties 
of the tails as a function of the SFE and $\tau_{\rm M}$. 
A comparison of the modelled tails with the observed ones will provide an independent constraint of the possible value of 
these fundamental parameters in the cluster formation theory.
We expect that the Gaia mission will reveal many tails associated with Galactic star clusters or confirm their absence; 
both results will be valuable at constraining the parameters. 
In our modelling, we aim particularly at the Pleiades due to their proximity and age; however our results 
can be easily generalised to other star clusters. 
Throughout this work, the cluster is assumed to orbit the Galaxy in its disk midplane and with zero eccentricity. 

It turns out that the morphology and kinematics of tail I and tail II are distinct. 
Unlike the properties of tail II, which are reasonably well understood \citep{Kupper2008,Kupper2010}, 
the properties of tail I have not been studied before. 
Thus, the morphology and kinematics of tail I is addressed in this paper.
First, we develop a (semi-)analytical model for some of the most important properties of 
tail I, and then we compare the model with direct N-body simulations. 
The N-body simulations in this paper represent only one choice of the SFE and $\tau_{\rm M}$, 
but their purpose is to justify the approximations undertaken in deriving the (semi-)analytical model. 

In paper II, we investigate the influence of the SFE and $\tau_{\rm M}$ on the formation and evolution of the tidal tails. 
We also study the cluster mass as another independent parameter.  
Taking into account the observed properties of the Pleiades cluster, we predict the physical extent and number of 
stars in the tidal tails for different values of the SFE and $\tau_{\rm M}$. 
For this prediction, we take into account Gaia accuracy as well as the contamination due to field stars. 
We also make use of the semi-analytic model of paper I to extend the models to higher mass clusters than the Pleiades.



\section{Semi-analytic model of the gas expulsion tidal tail}

\label{stailAnal}


\subsection{Underlying assumptions and starting equations}

Unlike tail II, which is gradually formed as stars evaporate from the cluster over hundreds 
of Myr, tail I is formed quickly as stars escape the cluster immediately after the gas expulsion event. 
Accordingly, we assume that all tail I stars escape at time $t=0$. 

Another difference between tails I and II is the magnitude of velocity of the escaping stars, which determines the allowed  
directions of escape. 
As found by \citet{Kupper2010}, the majority of stars forming tail II gets a velocity only slightly 
larger than the velocity needed to escape from the vicinity of Lagrange points L1 and L2 (of the system cluster-Galaxy), but 
the velocity is smaller that what is required for an escape in other directions. 
The preferential escape of stars near points L1 and L2 results in the typical S-shape of tail II.
Therefore, the speeds of escaping stars are of the order of the difference $\Delta C_{\rm J}$ between the maximum and minimum of the 
Jacobi potential evaluated at the cluster tidal radius (i.e. at $|\mathbf{r}| = r_{\rm t}$) in the orbital plane. 
It is straightforward to show that
\begin{equation}
\Delta C_{\rm J} = \frac{1}{2} \nu^2 \left( \frac{G M_{\rm cl}}{4 \omega^2 - \kappa^2} \right)^{2/3}, 
\label{eJacobiDiff}
\end{equation}
where $G$ and $M_{\rm cl}$ is the constant of gravity and the cluster mass, and $\omega$, $\kappa$ and $\nu$ is 
the orbital, epicyclic and vertical frequency of the Galaxy, respectively. 
This equation provides an escape speed of $\widetilde{v}_{\rm e,II} \approx \sqrt{2\Delta C_{\rm J}} = 2.2 \Kms$ for a $4.4 \times 10^3 \Msun$ cluster orbiting the 
Galactic potential of \citet{Allen1991} at the Galactocentric distance of $8.5 \Kpc$. 
In contrast, the escape speeds $\widetilde{v}_{\rm e,I}$ of stars forming tail I are of the order of the initial velocity 
dispersion $\sigma_{\rm cl}$ in the embedded star cluster, which is $\sigma_{\rm cl} = \sqrt{G M_{\rm cl}/(2 \; \sfe \; R_{\rm V})}$, 
where $R_{\rm V}$ is the cluster virial radius before gas expulsion. 
A $4.4 \times 10^3 \Msun$ Plummer cluster of $a_{\rm cl} = 0.23 \Pc$ and $\sfe = 1/3$ has $\sigma_{\rm cl} = 8.5 \Kms$, which is significantly 
larger than its $\widetilde{v}_{\rm e,II}$
\footnote{
Although the particular value of $\sigma_{\rm cl} = 8.5 \Kms$ might seem substantially higher than what is observed in young star-forming regions 
(e.g. \citealt{Kuhn2019}), 
we adopt such an example to have a testbed for the semi-analytic model with as high $\sigma_{\rm cl}/\widetilde{v}_{\rm e,II}$ as possible. 
On the other hand, this model is not unreasonable because it naturally follows for an initially virialised cluster with $a_{\rm cl} = 0.23 \Pc$. 
After rapid gas expulsion, the velocity dispersion of the remaining cluster decreases by a factor of several, bringing it closer to the values 
observed in young exposed clusters \citep{Rochau2010,Henault-Brunet2012}. 
Moreover, the majority of observed subclusters in star-forming regions is less massive, with a typical mass $M_{\rm cl} \approx 300 \Msun$ \citep{Kuhn2015}. 
A cluster of $300 \Msun$ and $a_{\rm cl} = 0.16 \Pc$ has $\sigma_{\rm cl} = 2.7 \Kms$, which is still larger than its $\widetilde{v}_{\rm e,II} = 0.9 \Kms$. 
Also note that $\sigma_{\rm cl}$ in the present paper is the 3D velocity dispersion, which is $\approx 1.7 \times$ higher than the 
observationally accessible 1D velocity dispersion.

}
.

The inequality $\widetilde{v}_{\rm e,II} \lesssim \widetilde{v}_{\rm e,I} \approx \sigma_{\rm cl}$ 
holds for any cluster with $R_{\rm V} \lesssim 2 \Pc$ and $M_{\rm cl} \gtrsim 200 \Msun$, 
so it is likely that it is fulfilled for the majority of star clusters in the current Galactic disc. 
The inequality implies that the faster stars from tail I are not sensitive to the potential
difference at the tidal radius, and they escape the cluster in any direction with the same ease.
Thus, in the semi-analytic model, we assume that stars escape isotropically. 
The validity of the approximation near the threshold $\sigma_{\rm cl} \approx \widetilde{v}_{\rm e,II}$ is explored in \refs{ssSmallerSigma}.


\iffigs
\begin{figure}
\includegraphics[width=\columnwidth]{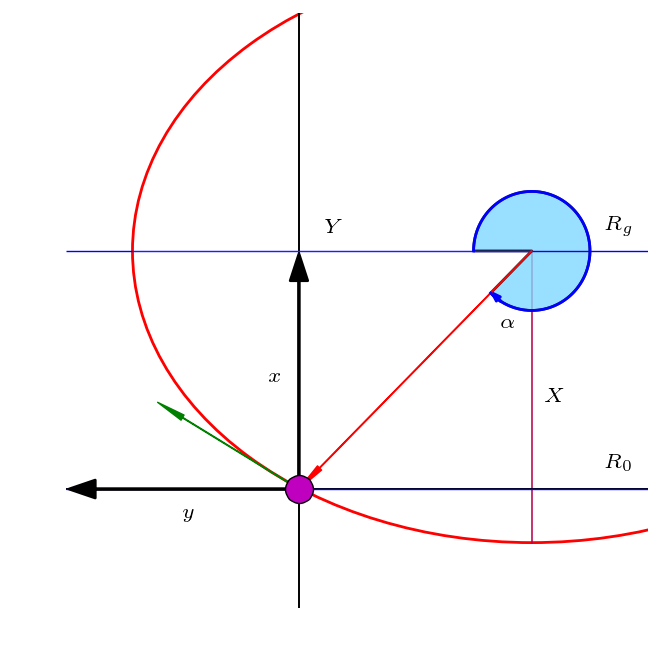}
\caption{Coordinate frame adopted for deriving the semi-analytic model of tail I. 
The frame is centred at the star cluster (the magenta circle), and the cluster remains at rest 
in the frame (the frame orbits the Galaxy). 
The coordinate axes $x$, $y$, as indicated by thick black arrows, follow 
the usual convention where the axis $x$ points towards the galactic anticentre and the axis $y$ points in the direction of the orbit. 
The cluster orbits the Galactic centre at radius $R_0$, which corresponds to $x = 0$ (thick blue line). 
The green arrow represents the initial trajectory of a star escaping the cluster. 
After escaping the cluster, the star moves on the red epicycle of semi-minor axis $X$, semi-major axis $Y$, 
and with guiding centre orbiting at Galactic radius $R_{\rm g}$, 
which corresponds to $x = X \sin \alpha$ in the adopted coordinates (thin blue line). 
The meaning of the escape angle $\alpha$ is shown by the blue curved arrow.}
\label{fConfig}
\end{figure} \else \fi


To derive the relevant equations, 
we adopt the coordinate frame to corotate with the cluster 
so that the cluster mass centre is at rest in this frame, and it lies at the origin of the coordinate axes (Fig. \ref{fConfig}).
Axes $x$ and $y$ point to the Galactic anticentre and in the direction of Galactic rotation, respectively. 
The axis $z$ forms a right-handed system with axes $x$ and $y$. 
The cluster orbits the galaxy on a circular orbit of radius $R_0$ and velocity $v_0$, so 
the radius of the guiding centre of the cluster is thus $R_0$. 

Consider a star escaping from the cluster at velocity with components ($v_{\rm R}$, $v_{\phi}$, $v_{\rm z}$) 
in the direction ($x$, $y$, $z$). 
The star then orbits its new guiding centre $R_{\rm g}$ on an ellipse with semi-major and semi-minor axes $Y$ and $X$, respectively, 
so its trajectory can be written as \citep{Binney2008}
\begin{subequations}
\begin{align}
x & = X\sin(\kappa t + \alpha) - X\sin(\alpha), \\
y & = Y\cos(\kappa t + \alpha) - Y\cos(\alpha) + v_{\rm g} t, \\
z & = \frac{\sqrt{v_{\rm R}^2 + v_{\phi}^2 + v_{\rm z}^2} \cos(\theta)}{\nu} \sin(\nu t + \zeta), \label{eeOrbitZ}
\end{align}
\label{eOrbit}
\end{subequations}
\noindent
where the angles $\alpha$ and $\theta$ denote the direction in which the star escaped the cluster ($\theta$ is 
the usual spherical angle measured from the $z$ axis and the meaning of angle $\alpha$ is shown in Fig. \ref{fConfig}), 
$v_{\rm g}$ is the relative velocity of the guiding centre at $R_{\rm g}$ 
to the guiding centre at $R_{\rm 0}$, and $\nu \equiv \pderrow{^2 \Phi(R,z)}{z^2}$. 
We set $\zeta = 0$ as all stars have $z = 0$ at $t = 0$.
Note that (see \citealt{Binney2008})
\begin{equation}
Y \equiv \gamma X = \frac{2 \omega}{\kappa} X.
\label{ey}
\end{equation}

Next, we evaluate $v_{\rm g}$ and $X$ as functions of the velocity components and escape direction of the escaping star.
At the time of leaving the cluster, the star moves at velocity $v_0+v_{\phi}$ around the galaxy. 
Subtracting the expressions for circular orbits,
\begin{equation}
\left. \pder{\Phi(R,z)}{R}\right|_{R_{\rm g}} - \left. \pder{\Phi(R,z)}{R}\right|_{R_0} = (v_0+v_{\phi})^2/R_{\rm g} - v_0^2/R_0,
\label{eExpansion}
\end{equation}
and using the expansion to the first order in small quantities $v_{\phi}$ and $R_{\rm g} - R_0$, so that 
$\pderrow{\Phi(R,z)}{R}|_{R_{\rm g}} - \pderrow{\Phi(R,z)}{R}|_{R_0} = \pderrow{^2 \Phi(R,z)}{R^2}|_{R_0} (R_{\rm g} - R_0)$, 
and $\pderrow{^2 \Phi(R,z)}{R^2}|_{R_0} = \kappa^2 - 3 \omega^2$, we arrive at
\begin{equation}
R_{\rm g} - R_0 = \frac{2 v_{\phi} \omega}{\kappa^2}.
\label{eR1}
\end{equation}
The velocity of the star's guiding centre $v_{\rm g}$ on its new orbit, relative to the cluster is given by
\begin{equation}
\left. v_{\rm g} = R_0 \pder{\omega}{R}\right|_{R_0} (R_{\rm g} - R_0),
\label{evel}
\end{equation}
where
\begin{equation}
\left. R_0 \pder{\omega}{R} \right|_{R_0} = \frac{\kappa^2}{2 \omega} - 2 \omega.
\label{evelExp}
\end{equation}

From Fig. \ref{fConfig} and \eq{eR1}, it follows for the semi-minor axis,
\begin{equation}
X = \frac{v_R}{\kappa \cos(\alpha)}.
\label{ex}
\end{equation}
The relationship between the velocity components and the angle $\alpha$, which apparently is
\begin{equation}
\frac{v_{\phi}}{v_{\rm R}} = -\frac{\kappa}{2 \omega} \tan(\alpha), 
\label{ealpha}
\end{equation}
enables us to express $X$ only as a function of $v_{\phi}$ and $v_{\rm R}$,
\begin{equation}
X = \frac{\sqrt{v_R^2 + 4\omega^2 v_{\phi}^2/\kappa^2}}{\kappa}.
\label{exv}
\end{equation}

\subsection{Evolution of the tail morphology and velocity structure}

\label{ssMorphol}


Several important tail properties can be obtained by substituting eqs. (\ref{ey}), (\ref{evel}), (\ref{ex}) and (\ref{exv}) 
into eqs. (\ref{eOrbit}). 
The time when the tail thickness in direction $x$ is zero occurs when $x/y$ is a constant for stars originating with any 
$v_{\phi}$ and $\alpha$. 
Using the equations above, one obtains 
\begin{equation}
\frac{x}{y} = \frac{x(\kappa, \alpha, t)}{y(\kappa, \alpha, t)},
\label{eTailThickness0}
\end{equation}
where
\begin{eqnarray}
x(\kappa, \alpha, t) & = & - \sin(\kappa t)\cos(\alpha) - \sin(\alpha) (\cos(\kappa t) - 1), \\ \nonumber
y(\kappa, \alpha, t) & = & \gamma (\sin(\kappa t)\sin(\alpha) + \cos(\alpha) (1 - \cos(\kappa t)) + \\ \nonumber
&& \left( \frac{\kappa^2}{2 \omega} - 2 \omega \right) \sin(\alpha) t/\gamma).
\label{eTailThickness05}
\end{eqnarray}
At general time $t$, \eq{eTailThickness0} provides different values of $x/y$ for stars escaping under different angles 
$\alpha$. 
However, at some instants, the tail thickness drops to zero. 
These events occur for time $t$ when \eq{eTailThickness0} is independent of $\alpha$, i.e. all stars
happen to lie on the same line regardless of the direction from which they escaped from the cluster. 
This condition is fulfilled when the terms $\sin(\alpha)$ and $\cos(\alpha)$ from the numerator and denominator 
are multiplies with the same constant $C$, i.e.
\begin{eqnarray}
\sin(\kappa t) & = & C(\cos(\kappa t) - 1), \\ \nonumber
1 - \cos(\kappa t) &=& C(\sin(\kappa t) + \frac{t}{\gamma} (\frac{\kappa^2}{2 \omega} - 2 \omega)),
\label{eTailThickness1}
\end{eqnarray}
which leads to the condition for zero tail thickness as
\begin{eqnarray}
(1 - \cos(\kappa t))^2  + && \\ \nonumber 
\sin(\kappa t) \left\{ \sin(\kappa t) + \frac{t}{\gamma} \left( \frac{\kappa^2}{2 \omega} - 2 \omega \right) \right\} & = & 0.
\label{eTailThickness2}
\end{eqnarray}
The condition has two types of solutions.
The first occurs when $\kappa t = 2\pi l$ for any integer $l$. 
In this case, the tail is aligned with axis $y$, i.e. $x/y = 0$. 
The other type of solutions do not occur in regular intervals, and we can find them only numerically. 
In these events, the value of $x/y$ is negative, 
which means that the leading part of the tail points 
from the direction of the rotation slightly towards the direction to the Galactic centre. 
The tilt of the tail relative to the axis $y$ at the next minimum thickness is always smaller 
than that at the previous minimum, so the tilt decreases with time (see Table \ref{tZeros}), 
and the tail becomes more aligned with the axis $y$. 

The two types of solutions alternate with the first type (analytical) solution having $x/y = 0$, followed by 
the second type (numerical) solution having $x/y$ of a small negative value. 
If we count the consecutive zeros of the tail thickness in time during the tail evolution, 
the first type solutions are odd, and second type solutions are even. 

The important property of both solutions is that they solely depend on the local 
galactic frequencies; the orbital frequency $\omega$ and the epicycle frequency $\kappa$, and they 
do not depend on any cluster property (e.g. the initial velocity dispersion or mass). 
The solutions as calculated for the local values of $\omega$ and $\kappa$ for the potential given by \eq{ePotTot} 
(i.e. at radius $R_0 = 8.5 \Kpc$, where $\omega = 8.38 \times 10^{-16} \Si$, $\kappa = 1.185 \times 10^{-15} \Si$) 
are listed in Table \ref{tZeros} for all events of zero tail thickness occurring in the first Gyr after the cluster formed.

\begin{table}
\begin{tabular}{cccc}
$t_{x0} [\Myr]$ & $-x/y$ & $t_{\sigma,x} [\Myr]$ & $t_{\sigma,y} [\Myr]$  \\
\hline
168.0 & 0.00 & 114.3 & 168.0 \\
228.7 & 0.33 & 203.2 & 234.5 \\
336.1 & 0.00 & 289.2 & 336.1 \\
406.4 & 0.19 & 374.3 & 409.8 \\
504.2 & 0.00 & 459.1 & 504.2 \\
578.4 & 0.13 & 543.6 & 580.9 \\
672.3 & 0.00 & 628.0 & 672.3 \\
748.7 & 0.10 & 712.3 & 750.6 \\
840.3 & 0.00 & 796.5 & 840.3 \\
918.1 & 0.08 & 880.7 & 919.7
\end{tabular}
\caption{List of time events when tail I thickness becomes zero ($t_{x0}$; first column), and 
when tail I velocity dispersion in directions $x$ and $y$ becomes zero ($t_{\sigma,x}$ and $t_{\sigma,y}$, respectively; 
third and fourth column). 
The tilt of the tidal tail $x/y$ at the instant of zero tail thickness is written in the second column. 
The table is calculated for frequencies $\omega$ and $\kappa$ corresponding to 
the potential of \eq{ePotTot} at Galactocenric radius $R_0 = 8.5 \Kpc$ 
(i.e. $\omega = 8.38 \times 10^{-16} \Si$, $\kappa = 1.185 \times 10^{-15} \Si$, $\nu = 2.92 \times 10^{-15} \Si$), 
and it lists all the events occurring in the first Gyr after the formation of the cluster.}
\label{tZeros}
\end{table}

Similarly, taking the time derivative of \eq{eOrbit} and substituting in the same equations, we can express the velocity 
components $\dot{x}$ and $\dot{y}$ in directions $x$ and $y$ as functions of the distance $y$ along the tail. 
There are also events when the velocity $\dot{x}$ is constant at a given distance $y$, which means that 
all stars at a given $y$ move at the same speed $v_x$, i.e. the $x-$ component of the velocity dispersion $\sigma$ is zero. 
The condition for this reads
\begin{equation}
\sin(\kappa t)(\cos(\kappa t) - 1) = \cos(\kappa t) \left\{\sin(\kappa t) + \kappa t\frac{\kappa^2 - 4 \omega^2}{4 \omega^2} \right\}.
\label{eTailSigmaY2}
\end{equation}
These events are listed in the third column of Table \ref{tZeros}, and 
unlike the minima of $x/y$ or $\sigma_{\rm y}$ (see below), they do not admit the analytical solution $\kappa t = 2\pi l$, 
so they do not coincide with any minima of $x/y$ or $\sigma_{\rm y}$. 

Likewise, the velocity dispersion $\sigma_{\rm y}$ in the direction $y$ reaches zero when
\begin{align}
& (1 - \cos(\kappa t))(1 - \cos(\kappa t) - \frac{\kappa^2}{4 \omega^2}) + \nonumber \\
& \sin(\kappa t) \left\{ \sin(\kappa t) + \frac{t}{\gamma} \left( \frac{\kappa^2}{2 \omega} - 2 \omega \right) \right\} = 0.
\label{eTailSigmaX2}
\end{align}
Similar to \eq{eTailThickness2}, this equation has an analytic solution for $\kappa t = 2\pi l$ for any integer $l$, 
so the zeros of velocity dispersion $\sigma_{\rm y}$ coincide with minima in the tail thickness for tails aligned with $x/y = 0$. 
There is another type of solutions, which must be found numerically, however these events do not exactly 
coincide with the tail thickness minima, but they occur several Myr later (see the fourth column of Table \ref{tZeros} 
for all events of zero $\sigma_{\rm y}$). 

As with the tail thickness, the minima of $\sigma_{\rm x}$ and $\sigma_{\rm y}$ depend only on the 
Galaxy parameters $\omega$ and $\kappa$ and not on the star cluster properties.

The $y-$ component of the bulk velocity of the tail can be expressed analytically at times 
$t_{\rm y,b} = 2\pi l/\kappa $ as
\begin{equation}
\dot{y}(y) = \frac{y}{t_{\rm y,b}} \frac{\kappa^2}{\kappa^2 - 4 \omega^2},
\label{eTailSigmaX}
\end{equation}
which means that at these time events, the bulk velocity of the tail linearly increases along the tail, 
and that at a given position $y$ in the tail, the bulk velocity decreases with time as $t_{\rm y,b}^{-1}$.

After deriving the formulae for tail kinematics in the plane $xy$, which are new results, we 
list the well known formulae for the vertical motion. 
The motion in the $z$ direction is particularly simple as the Hamiltonian of a stellar orbit in an axially symmetric galaxy 
for nearly circular orbits close to its plane ($z \lesssim 300 \Pc$) 
can be separated between the radial motion and the vertical motion \citep{Binney2008}, 
which leads to \eq{eeOrbitZ}. 
We verify that the condition is fulfilled for the stars released due to gas expulsion.
The vertical velocity is $\dot{z} = \sqrt{v_{\rm R}^2 + v_{\phi}^2 + v_{\rm z}^2} \cos(\theta)$ \eqp{eeOrbitZ}, 
and the velocities of escaping stars 
from clusters of mass $\lesssim 10^4 \Msun$ are $v_{\rm e,I} \lesssim 10 \Kms$, which together 
with the value of $\nu$ of the potential of \eq{ePotTot}, which is $\nu = 2.92 \times 10^{-15} \; \rm{s}^{-1}$, 
yields the maximum displacement of $z \approx 110 \Pc$. 

Although the expansion of tail I and its oscillations in the $xy$ plane are quite complex, 
the tail motion in the $z$ direction is simple; the thickness in the $z$ direction of tail I harmonically oscillates 
with frequency $\nu$, so that its $z$ thickness reaches zero at time $t = \pi l/\nu$ for any non-negative integer $l$.

\subsection{Evolution of tail half-mass radius}


Although the semi-analytic model introduced in the previous section 
estimates the time of minima for different important quantities of tail I, 
it does not estimate the extent of tail I. 
To study the size of the tail, we adopt the concept of Lagrangian radii for the tail, i.e. the radii 
centred on the cluster which enclose a specific fraction of the total tail mass. 
In the semi-analytic study, we focus on the half-mass radius, i.e. the radius enclosing 
$1/2$ of the total tail mass. 
Naturally, we exclude the cluster itself (defined as being composed of all stars  
gravitationally bound to the cluster) for the purpose of these calculations.
The numerical simulations presented in Paper II show that the mass function (MF) of tail I is independent of position 
(assuming the cluster is not mass segregated when the gas is expelled), so 
the half-mass radius for any group of stars, and thus half-number radius for any group of stars 
are close to each other regardless of the stellar spectral type.

\iffigs
\begin{figure}
\includegraphics[width=\columnwidth]{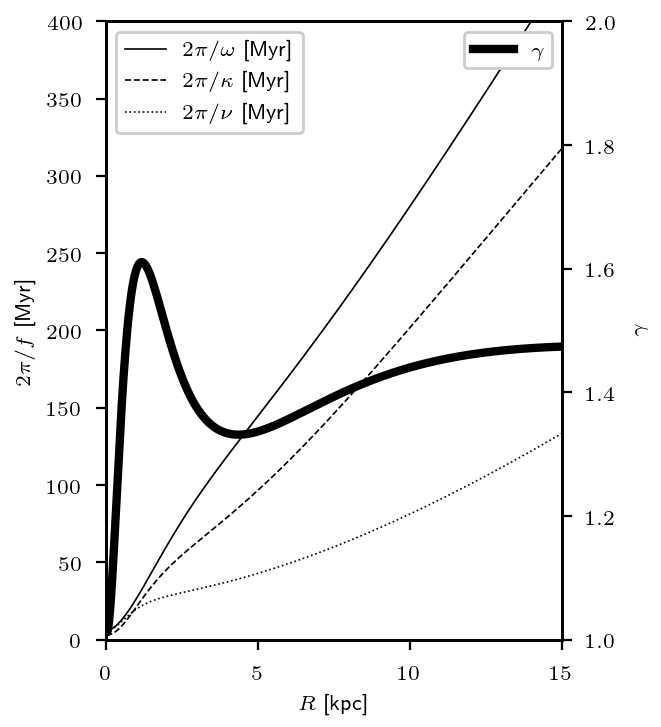}
\caption{Timescales corresponding to the Galactic frequencies $f$: $\omega$, $\kappa$ and $\nu$ (left axis), 
and the ratio $\gamma = 2\omega/\kappa$ (right axis) as a function of Galactocentric radius $R$ 
for the \citet{Allen1991} Galaxy model \eqp{ePotTot}.}
\label{fGalModel}
\end{figure} \else \fi

\iffigs
\begin{figure}
\includegraphics[width=\columnwidth]{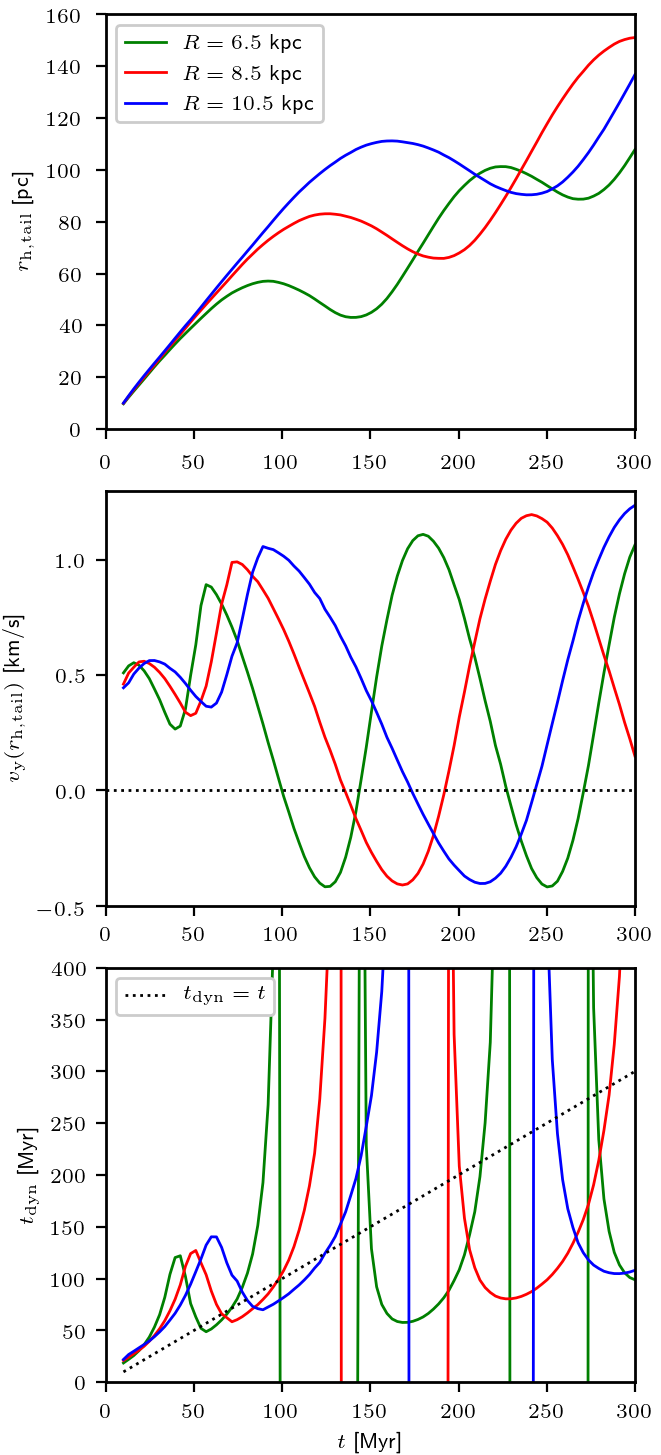}
\caption{\figpan{Top panel:} Time dependence of the tail half-mass radius $r_{\rm h,tail}$ for 
the semi-analytic model \eqp{erhsemiAnl}.
\figpan{Middle panel:} Time evolution of the mean velocity $v_{\rm y}(r_{\rm h,tail})$ near the radius $r_{\rm h,tail}$. 
The velocity oscillates substantially and it becomes negative when the tail moves towards the cluster. 
\figpan{Lower panel:} Time evolution of the dynamical age of the tail, i.e. $t_{\rm dyn} = r_{\rm h,tail}/v_{\rm y}(r_{\rm h,tail})$. 
The black dotted line represents the real age $t_{\rm dyn} = t$. 
Note that the dynamical age is often substantially different from the real age, 
making $t_{\rm dyn}$ an unreliable estimate for the real age. 
All plots are calculated for clusters orbiting at three Galactocentric radii ($6.5 \Kpc$, green line; 
$8.5 \Kpc$, red line; $10.5 \Kpc$, blue line), and for stars escaping at velocity $v_{e} = 1 \Kms$.}
\label{frh_semiAnl}
\end{figure} \else \fi

The results of Sect. \ref{ssMorphol} do not put any constraint on the magnitude of 
the velocity of escaping stars, $v_{\rm e} = \sqrt{v_{\rm R}^2 + v_{\phi}^2 + v_{\rm z}^2}$, 
so these results are valid for stars which escape the cluster with a range of velocities. 
To simplify the problem here, we further assume that all stars escape the cluster at the same velocity $v_{\rm e}$, but 
still in different directions as specified by angles $\alpha$ and $\theta$. 

With this definition of $v_{\rm e}$, \eq{evel} and \eq{exv} read
\begin{equation}
v_{\rm g} = -\frac{v_{\rm e} \tan(\alpha)\mathrm{sign}(\cos(\alpha))(1 - \gamma^2)|\sin(\theta)|}{\gamma \sqrt{1 + \tan^2(\alpha)/\gamma^2}},
\label{evgve}
\end{equation}
\begin{equation}
X = \frac{v_{\rm e} |\sin(\theta)|}{\kappa |\cos(\alpha)| \sqrt{1 + \tan^2(\alpha)/\gamma^2}},
\label{exvve}
\end{equation}
which together with eqs. (\ref{eOrbit}) and given orbital frequencies $\omega$, $\kappa$ and $\nu$, 
enable us to express the distance $r(t, v_{\rm e}, \alpha, \theta, \gamma, \kappa, \nu) \equiv \sqrt{x^2 + y^2 + z^2}$ at any time $t$. 
The dependence of the timescales for the frequencies $\omega$, $\kappa$, $\nu$ and the ratio $\gamma = 2\omega/\kappa$ 
for the potential of \eq{ePotTot} is shown in Fig. \ref{fGalModel}. 
All the three frequencies decrease monotonically with Galactocentric radius $R$. 
The ratio $\gamma$ attains the value of $1.415$ near the Solar radius (at $8.5 \Kpc$).

The half-mass radius of the tail is the number $r_{\rm h,tail}$ which fulfils 
\begin{align}
& \int_0^{\pi/2} \int_0^{2\pi} \min(r(t, v_{\rm e}, \alpha, \theta, \gamma, \kappa, \nu), r_{\rm h,tail}) \sin(\theta) \dd \theta \dd \alpha \nonumber \\
& = \frac{1}{2} \int_0^{\pi/2} \int_0^{2\pi} r(t, v_{\rm e}, \alpha, \theta, \gamma, \kappa, \nu) \sin(\theta) \dd \theta \dd \alpha.
\label{erhsemiAnl}
\end{align}
Since the dependence of $r(t, v_{\rm e}, \alpha, \theta, \gamma, \kappa, \nu)$ on velocity $v_{\rm e}$ is particularly simple, i.e. 
\begin{equation}
r(t, v_{\rm e}, \alpha, \theta, \gamma, \kappa, \nu) = v_{\rm e} r'(t,\alpha, \theta, \gamma, \kappa, \nu),
\label{erprime}
\end{equation}
the value of $r_{\rm h,tail}$ is only a function of the Galactic potential multiplied by velocity $v_{\rm e}$. 
This means, that time evolution of $r_{\rm h,tail}$ for any $\tilde{v}_{\rm e}$ can be trivially obtained by multiplying 
the evolution of $r_{\rm h,tail}$ calculated for one fixed value of $v_{\rm e}$ by $\tilde{v}_{\rm e}/v_{\rm e}$. 

The dependence of $r_{\rm h,tail} = r_{\rm h,tail}(t)$ as calculated 
by \eq{erhsemiAnl} for $v_{\rm e} = 1\Kms$ at Galactocentric radii $6.5 \Kpc$, $8.5 \Kpc$ and $10.5 \Kpc$ is shown 
in the upper panel of Fig. \ref{frh_semiAnl}.
The extent of the tail increases non-monotonically, changing from expansion to contraction first at $t \approx 120 \Myr$ 
(for $R_0 = 8.5 \Kpc$), and then again starts expanding at $t \approx 180 \Myr$. 
Another tail contraction starts at $t \approx 295 \Myr$, so the periods of expansion are interspersed by periods 
of contraction. 

Note that the tail is stretched mainly in the $y$ direction (eqs. \ref{eOrbit}), 
so the contributions from $x$ and $z$ component 
becomes unimportant as the tail stretches with time. 
Thus, the half-mass radius $r_{\rm h,tail}$ is very close to the half-mass radius in the direction $y$. 

The velocity near the half-mass radius $v_{\rm y}(r_{\rm h,tail})$  for 
$v_{\rm e} = 1 \Kms$ is plotted in the middle panel of Fig. \ref{frh_semiAnl}, 
showing oscillations where longer intervals of tail expansion (when $v_{\rm y}(r_{\rm h,tail}) > 0$) 
are interspersed with shorter intervals of tail contraction ($v_{\rm y}(r_{\rm h,tail}) < 0$). 
This pattern is present at all the Galactocentric radii, where the value of $v_{\rm y}(r_{\rm h,tail})$ 
for clusters located closer to the inner Galaxy oscillates with a shorter period than that in the outer Galaxy.
Due to the structure of \eq{erprime}, the velocity $v_{\rm y}(r_{\rm h,tail})$ for any cluster with characteristic 
escape speed $\tilde{v}_{\rm e}$ can be obtained by multiplying $v_{\rm y}(r_{\rm h,tail})$ by $\tilde{v}_{\rm e}/1 \Kms$. 

The lower panel of Fig. \ref{frh_semiAnl} compares  
the real cluster age (the dotted line) 
with the dynamical age of the tail $t_{\rm dyn}$, i.e. $t_{\rm dyn} = r_{\rm h,tail}/v_{\rm y}(r_{\rm h,tail})$ (solid lines). 
The dynamical age often differs by factor of $\sim 2$ from the real cluster age, 
and at the time when the tail starts contracting, the dynamical time suddenly drops and even becomes negative, 
demonstrating that the dynamical age is rather a poor estimate for the real age.

\section{Numerical methods and initial conditions}

\label{sInitCond}

\subsection{Numerical methods}

The present clusters are evolved by the code \nbdvi \citep{Aarseth1999,Aarseth2003}, 
which integrates trajectories of stars by the 4th order Hermite predictor-corrector scheme 
\citep{Makino1991} with adaptive block time-steps.
The force acting on each star is split into its regular and irregular part according to the 
Ahmad-Cohen method \citep{Ahmad1973,Makino1991}.
Close passages between stars and interactions in two- and more- stellar subsystems are treated by regularisation techniques 
\citep{Kustaanheimo1965,Aarseth1974a,Mikkola1990}. 
These sophisticated numerical methods result in a fast and very accurate code. 
Stellar evolution is treated by synthetic evolutionary tracks \citep{Hurley2000} for Solar metallicity. 

When a star escapes the cluster, its trajectory is integrated in the external potential of the Galaxy. 
For the Galactic potential, we adopt the model of \citet{Allen1991}, where the
Galaxy consists of three components: the central part, disc and halo. 
The potential of the central part and the disc is approximated by the Miyamoto-Nagai potential \citep{Miyamoto1975}, 
\begin{equation}
\phi_{i}(R,z) = - \frac{G M_i}{\sqrt{R^2 + (a_{\rm i} + \sqrt{b_{\rm i}^2 + z^2})^2}},
\label{ePotMN}
\end{equation}
where $i = 1$ for the central part and $i = 2$ for the disc. 
$G$ is the gravitational constant. 
The Galaxy is described in cylindrical coordinates $(R,z)$ with the origin at the Galactic centre. 
The central part of the Galaxy is described by parameters $M_1 = 1.41 \times 10^{10} \Msun$, $a_1 = 0$, $b_1 = 0.387 \Kpc$, 
so the model degenerates to the Plummer potential of scale-length $b_1$. 
The parameters of the disc are $M_2 = 8.56 \times 10^{10} \Msun$, $a_2 = 5.32 \Kpc$, $b_2 = 0.25 \Kpc$. 
The potential of the dark matter halo is given by
\begin{eqnarray}
\phi_{3}(R) & = & -\frac{G M(R)}{R} - \frac{G M_3}{1.02 a_3} \times \nonumber \\
&& \Bigg[ \left\{ \ln \left( 1 + (R_{\rm d}/a_3)^{1.02} \right) - 
\frac{1.02}{1 + (R_{\rm d}/a_3)^{1.02}} \right\} - \nonumber \\
&& \left\{ \ln \left( 1 + (R/a_3)^{1.02} \right) - 
\frac{1.02}{1 + (R/a_3)^{1.02}} \right\} \Bigg],
\label{ePotHalo}
\end{eqnarray}
where 
\begin{equation}
M(R) = \frac{M_3 (R/a_3)^{2.02}}{1 + (R/a_3)^{1.02}}, 
\label{eMRHalo}
\end{equation}
is the halo mass inside the sphere of radius $R$, $R_{\rm d}$ is a radius determining the constant in 
the halo potential (we adopt $R_{\rm d} = 100 \Kpc$), 
and $M_3 = 10.7 \times 10^{10} \Msun$, $a_3 = 12 \Kpc$. 
The halo may be viewed as a representation of the phantom dark matter halo \citep{Lughausen2015}.
The total potential of the Galaxy is a superposition of the three potentials, i.e.
\begin{equation}
\phi_{\rm Gal} = \phi_{1} + \phi_{2} + \phi_{3}.
\label{ePotTot}
\end{equation}

To approximate the gas expulsion from the embedded cluster, we use the model of \citet{Kroupa2001b}, where the 
gaseous potential is given by 
\begin{equation}
\phi_{\rm gas}(t) = -\frac{G M_{\rm gas}(t)}{\sqrt{a_{\rm gas}^2 + r^2}},
\label{ePotGas}
\end{equation}
where $r$ is the distance from the cluster centre, $a_{\rm gas}$ is the Plummer scale-length, and $M_{\rm gas}(t)$ the gas mass.
We choose $a_{\rm gas}$ to be identical to the Plummer scale-length $a_{\rm cl}$ for the stellar distribution. 
The removal of gas due to the action of newly formed stars is approximated by a decrease of the gas mass as 
\begin{equation}
M_{\rm gas}(t) = M_{\rm gas}(0) \exp{\{-(t - t_{\rm d})/\tau_{\rm M}\}}, \; t > t_{\rm d}
\label{ePotGaseous}
\end{equation}
where $M_{\rm gas}(0)$ is the initial mass of the gas within the cluster,
$t_{\rm d}$ the time delay after gas expulsion
commences, and $\tau_{\rm M}$ the time-scale of gas removal.

Following \citet{Kroupa2001b}, we adopt $t_{\rm d} = 0.6 \Myr$ to reflect the embedded phase before gas expulsion starts, 
which is about the lifetime of ultra compact HII regions \citep{Wood1989,Churchwell2002}. 
In this work, we adopt the usual choice of gas expulsion time-scale $\tau_{\rm M}$, which is motivated by the idea that
the gas is expelled at the velocity of the order of the sound speed in ionised hydrogen, i.e. $\tau_{\rm M} = a_{\rm gas}/(10 \Kms)$. 
This means that the gas is expelled on a time-scale shorter than the stellar half-mass crossing time $t_{\rm h}$ 
(impulsive gas removal), so
the stellar component adjusts only slightly during the gas expulsion, and is more affected by it. 
In paper II, we also explore models with $\tau_{\rm M} \gg t_{\rm h}$ (adiabatic gas removal). 

\subsection{Initial conditions}

At the beginning, the star cluster is a Plummer sphere of stellar mass $M_{\rm cl}$ and Plummer radius 
$a_{\rm cl}$ in virial equilibrium, and it is populated 
by stars by the method described by \citet{Aarseth1974b}. 
The stellar masses are randomly drawn from the canonical initial mass function 
in the form of (see \citealt{Kroupa2001a,Kroupa2013}),
\begin{equation}
\nder{N}{m} = 
\begin{cases}
c_1 m^{-1.3} \,\, 0.08 \Msun \leq m \leq 0.5 \Msun, \\
c_2 m^{-2.3} \,\, 0.5 \Msun < m, \\
\end{cases}
\label{eIMF}
\end{equation}
with the lower stellar mass limit being at the hydrogen burning limit $0.08 \Msun$, and the 
upper stellar mass limit being at $80 \Msun$, which is adopted 
according to the relation between the maximum mass of a star as a function 
of the mass of its host star cluster \citep{Weidner2010}, 
which we adopt to be $\approx 4400 \Msun$ (each cluster contains 10000 stars). 
Constants $c_1$ and $c_2$ are chosen so that the function $\nderrow{N}{m}$ is continuous at $0.5 \Msun$.
The metallicity is close to solar, $Z = 0.02$. 
Since this is the first attempt to study the phenomenon of two tails formed by a star cluster, 
we do not consider primordial binary stars for the sake of simplicity. 
The clusters are also not primordially mass segregated. 
The clusters are on circular orbits of speed $220 \Kms$, i.e. $(v_{\rm R}, v_{\rm \phi}, v_{\rm z}) = (0, 220, 0) \Kms$,  
at a Galactocentric distance $8.5 \Kpc$. 

\section{The structure and kinematics of tail I and tail II}

\label{sCompare}

\iffigs
\begin{figure*}[h!]
\includegraphics[width=\textwidth]{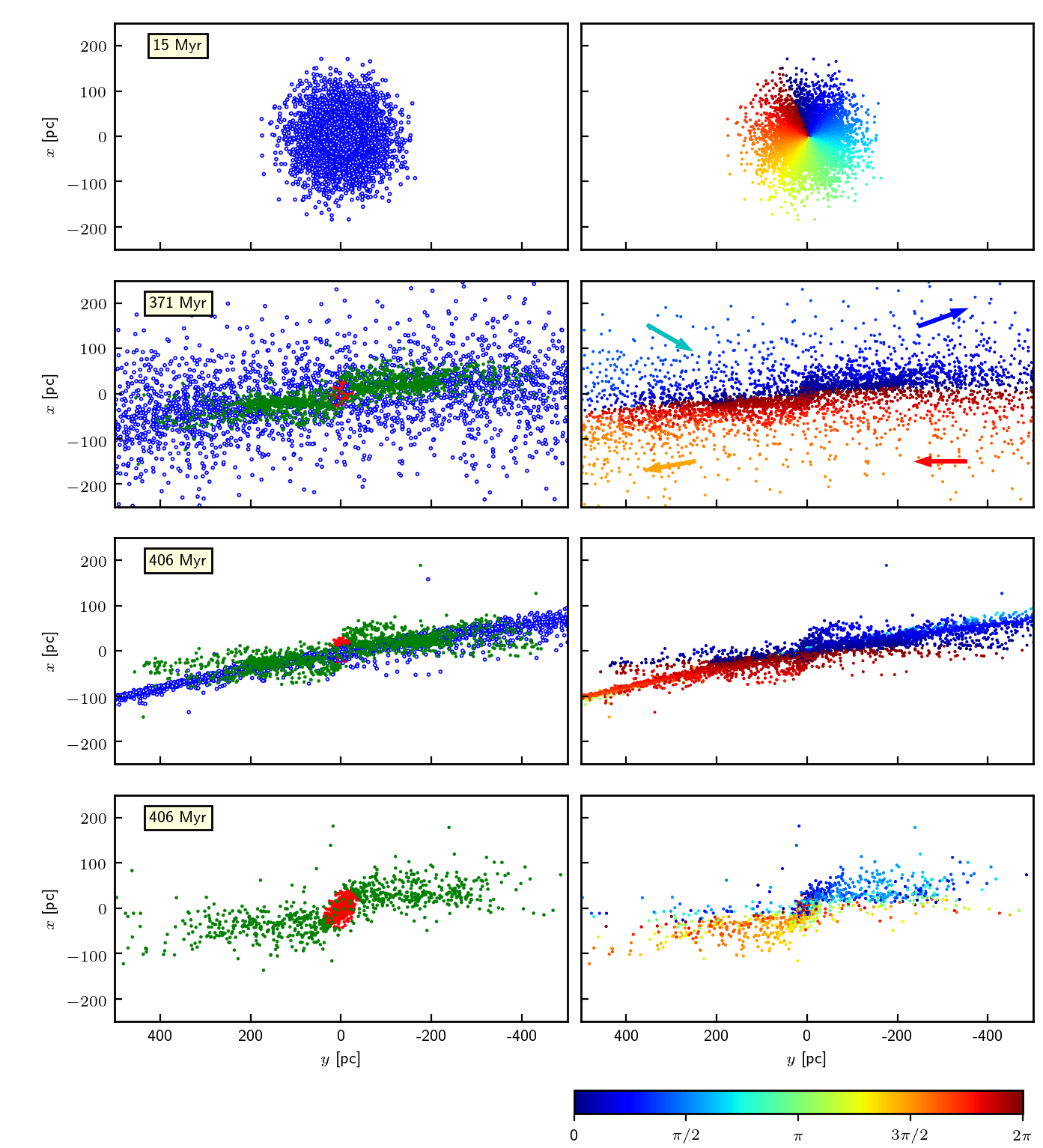}
\caption{Morphology of the tidal tails. From top to bottom: Model C10G13 at $15 \Myr$,
model C10G13 at $371 \Myr$ (maximum thickness of tail I),
model C10G13 at $406 \Myr$ (minimum thickness of tail I), and model C10W1 at $406 \Myr$.
Each star is represented by a coloured dot.
\figpan{Left column:} The structure of tail I (blue dots), and tail II (green dots).
Members of the star cluster are shown by red dots.
\figpan{Right column:} The direction of stellar motion in the tails.
The colour gives the direction of motion where the angle is measured clockwise from the unit vector pointing in the
direction of positive $x$ axis (see the arrows on the upper middle panel and the colourbar).
The cluster moves in the positive $y$ direction around the Galaxy.}
\label{fvfield_2columns}
\end{figure*} \else \fi

In order to highlight the difference between tails I and II, we here 
adopt rather extreme models of initial conditions for star clusters. 
These models are not necessarily meant to reproduce the observed clusters, but the 
main focus is on (i) describing the structural and kinematical differences between the two tails, 
and (ii) checking the validity of the semi-analytic model. 
Models including other choices of parameters are studied in Paper II.

The first model (hereafter C10G13) has $M_{\rm gas}(0) = 2 M_{\rm cl}$, i.e. the star formation efficiency is 33\%. 
The Plummer scale-length $a_{\rm cl}$ is $0.23 \Pc$, which is chosen according to the relation between the birth mass 
and the birth radius of star clusters suggested by \citet{Marks2012}. 
The other model (hereafter C10W1) is gas free, and its Plummer scale-length $a_{\rm cl}$ is $1 \Pc$. 
We set the parameter $a_{\rm cl}$ for the gas free model substantially larger than for the model C10G13 
to have comparable half-mass radii $r_{\rm h}$ for both clusters after the gas rich model expands and revirialises 
due to gas expulsion (cf. Fig. 6 in paper II). 

We define the stars belonging to tail I as the ones which escape the cluster before the time
\begin{equation}
t_{\rm ee} \equiv 2 r_{\rm t}/(0.1 \sigma_{\rm cl} (t = 0)),
\label{etee}
\end{equation}
where $r_{\rm t}$ is the tidal radius of the cluster.
The stars belonging to tail II escape after $t_{\rm ee}$.
The definition of $t_{\rm ee}$ is motivated by the fact that the stars released due to gas expulsion
escape at a velocity comparable to the initial cluster velocity dispersion $\sigma_{\rm cl} (t=0)$,
so tail I is composed of stars which travel the distance to the tidal radius in a fraction ($0.1$)
of the initial velocity dispersion.
The particular value of $0.1$ is implied from the shape of the velocity dispersion in the Plummer model,
where the vast majority of stars move at the velocity in excess of $0.1 \sigma_{\rm cl}$.
The results are insensitive to the particular choice of the numerical factor.
There are some stars which escape due to evaporation earlier than $t_{\rm ee}$, and stars released due to
the gas expulsion at speed smaller than $0.1 \sigma_{\rm cl} (t=0)$ so they escape after $t_{\rm ee}$, but the number of
these stars is small.

The shallowing of the potential due to gas expulsion in model C10G13 
unbinds 58\% of its stars by time $t_{\rm ee} = 30 \Myr$, the escape rate then rapidly decreases, and 
the cluster loses 68\% of its initial stellar content by $300 \Myr$. 
The evaporations and ejections alone in the gas free model C10W1 unbind substantially less stars, 
only 0.3\% and 5\% of the total cluster population by $30 \Myr$ and $300 \Myr$, respectively, 
so the gas expulsion model forms 
a tail which contains at least $10$ times more stars 
during the first $\approx 300 \Myr$ of evolution.

\iffigs
\begin{figure}
\includegraphics[width=\columnwidth]{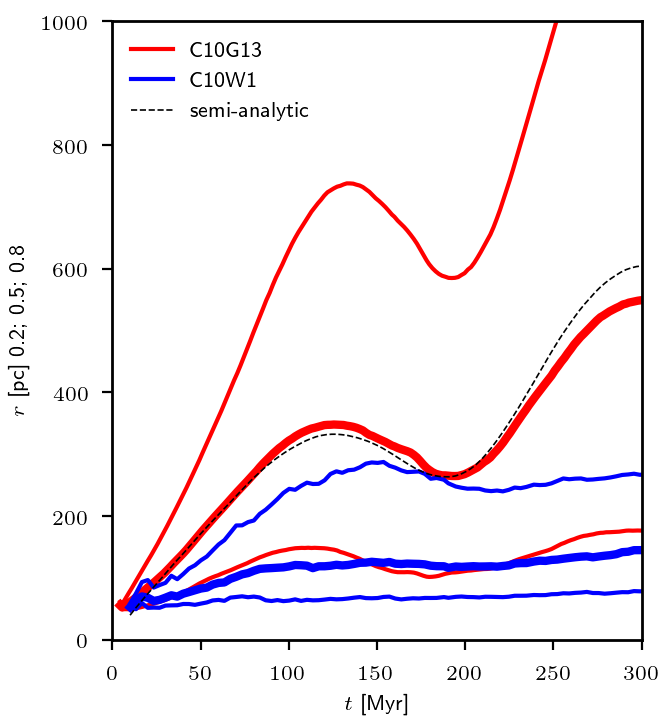}
\caption{Evolution of the 0.2, 0.5 and 0.8 Lagrangian radius of the tidal tail for the number of stars 
for model C10G13 (red lines) and model C10W1 (blue lines). 
The half-number radius is indicated by the thick lines. 
The dashed line represents the semi-analytical half-number radius calculated by \eq{erhsemiAnl} and scaled to $v_{\rm e} = 4 \Kms$, 
which is chosen according to the median $\widetilde{v}_{\rm e,I}$ for model C10G13. 
This demonstrates an excellent agreement of the N-body model C10G13 with the semi-analytic result of Sect. \ref{stailAnal}.}
\label{flagrtail}
\end{figure} \else \fi

\iffigs
\begin{figure*}
\includegraphics[width=\textwidth]{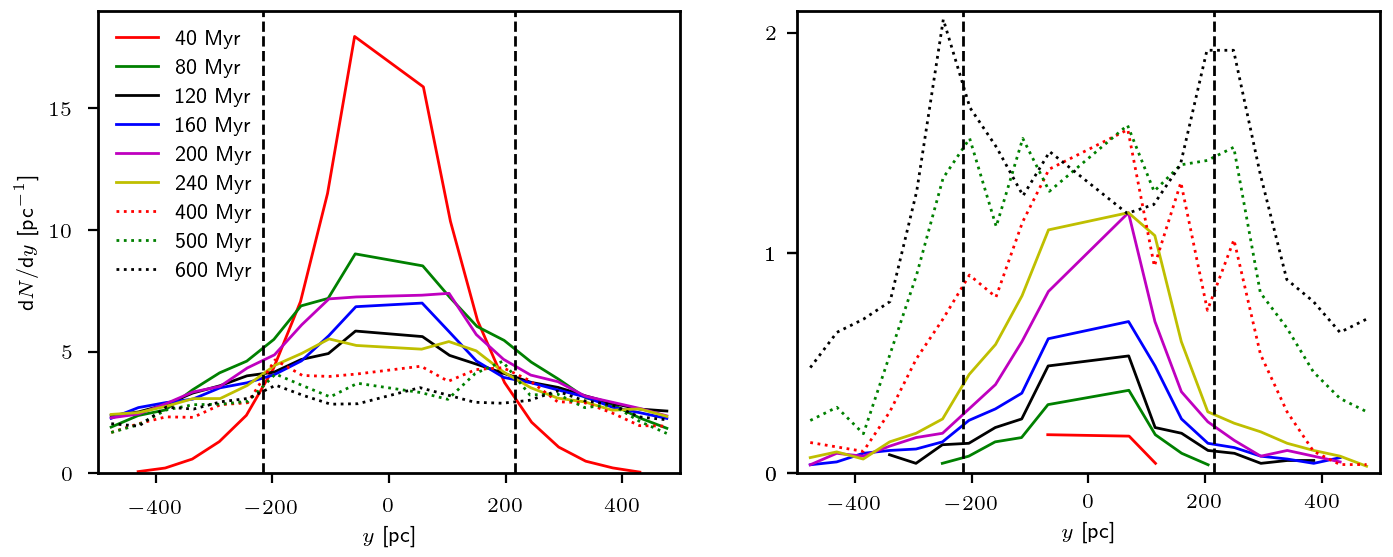}
\caption{Number density distribution of the tail as a function of the distance $y$ along the tail for model C10G13 (left panel)
and model C10W1 (right panel) at selected times as indicated by the lines.
Plots at times earlier than $300 \Myr$ (solid lines) do not indicate epicyclic overdensities. 
Epicyclic overdensities appear later ($400 - 600 \Myr$; dotted lines). 
The epicyclic overdensities are located at a distance very close to the 
one predicted by \citet{Kupper2008}, which is indicated by the vertical dashed lines.}
\label{fDensProf}
\end{figure*} \else \fi

\iffigs
\begin{figure*}[h!]
\includegraphics[width=\textwidth]{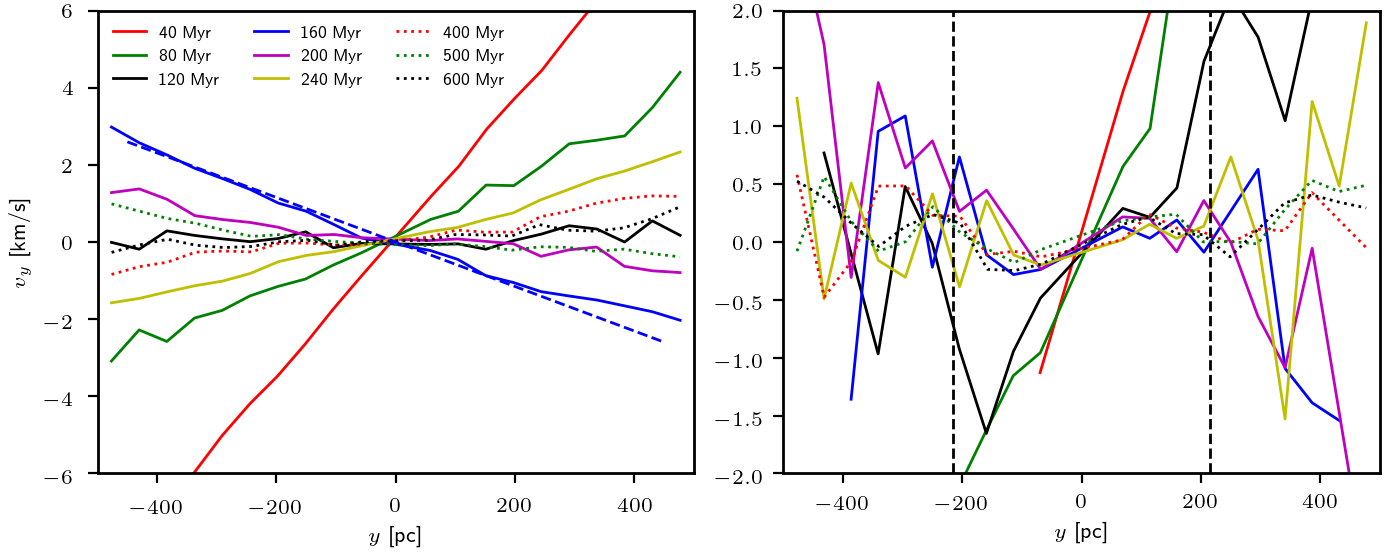}
\caption{Evolution of the bulk velocity $v_{\rm y}$ of the tidal tail along the direction $y$ at selected times as 
indicated by solid or dotted lines.
\figpan{Left panel:} Model C10G13. The blue dashed line indicates the analytical
solution given by \eq{eTailSigmaX} at $t = 168 \Myr$.
\figpan{Right panel:} Model C10W1. The vertical dashed lines represent the positions of $y_{\rm C}$ according to \citet{Kupper2008}. 
The epicyclic overdensities, which develop near the predicted positions $\pm 240 \Pc$ at $400 - 600 \Myr$, 
are places where stars slow down or even reverse their motion, i.e. move towards the cluster. 
}
\label{fvelalongx}
\end{figure*} \else \fi

\subsection{The morphology of the tails}

Figure \ref{fvfield_2columns} compares the morphology (left column) and velocity (right column) 
structure of the tidal tails of the two models. 
First, we study the morphology. 
Tails I and II are depicted as blue and green dots, respectively; the tail membership is based on \eq{etee}. 
The star cluster members are shown by red dots.

At the beginning of evolution of model C10G13 ($t = 15 \Myr$; upper row), 
the tail expands isotropically forming approximately a sphere. 
This means that the escaping stars are insensitive to the variation of the Jacobi potential around the radius $r_{\rm t}$, 
which is in agreement with the assumption in Sect. \ref{stailAnal} as the median escape velocity of the 
tail I stars ($\widetilde{v}_{\rm e,I} \approx 4 \Kms$) is substantially larger than the potential difference 
at the surface of the sphere of radius $r_{\rm t}$ (at $|\mathbf{r}| = r_{\rm t}$, which is the approximate 
location of the Lagrange points L1 and L2), which corresponds to a velocity of $\lesssim 1 \Kms$. 

The second implication of the high escape speed is the large epicycle radius $X$ \eqp{exv},
which exceeds the tidal radius $r_{\rm t}$ of the cluster
as the escape speed of $0.9 \Kms$ implies $X = 24 \Pc$, which corresponds to the value of $r_{\rm t}$. 
The large escape speed also results in a substantially longer tail at a given time.
Thus, the stars in tail I are brought to significantly larger or shorter Galactocentric radii than $R_0 \pm r_{\rm t}$
forming an extended envelope around the cluster.

To illustrate the time variability of tail I as found in \refs{stailAnal}, we plot the tail 
between its third and fourth predicted minimum of thickness (at $371 \Myr$; upper middle row of \reff{fvfield_2columns}), 
and at the fourth predicted minimum of thickness ($406 \Myr$; lower middle row). 
The tail thickness changes significantly as well as the tilt relative to the axis $x=0$. 
The volume density close to the axis $x=0$ also significantly increases near the time of the minimum thickness facilitating 
the detection of the tail to a larger distance from the cluster. 
The minimum of tail thickness in direction $x$ occurs almost at the time predicted in \reft{tZeros}, and also the tilt $x/y \approx -0.2$ is 
very close to the value predicted by the analytic formula.

%

In contrast, tail II arises gradually, with the evaporating rate nearly constant with time \citep{Baumgardt2003}, 
so apart from continuing expansion from the cluster, tail II shows no tilt or oscillations of thickness
(cf. the upper middle and lower middle panels). 
The escape speed is typically by a factor of $2-3$ smaller 
than the cluster velocity dispersion $\sigma_{\rm cl}$. 
Consequently, the majority of tail II stars escape near the Lagrange points L1 and L2, 
where the potential barrier is the shallowest, 
and the slow velocity of escaping stars results in epicycle semi-minor axes $X$ being usually not larger than the tidal radius. 
Thus, for example, the stars in the trailing tail (which were released from $x \approx r_{\rm t}$), move in the $x$ 
direction mainly from $x \approx 0$ to $x \approx 2 r_{\rm t}$, with few stars reaching $x < 0$. 
From symmetry, the opposite holds for the leading tail, which forms the characteristic S-shape of tail II. 
The morphology of tail II is almost identical to that of the tail caused by evaporating 
stars in the gas free model (cf. with model C10W1; lower row of \reff{fvfield_2columns}). 


\iffigs
\begin{figure*}
\includegraphics[width=\textwidth]{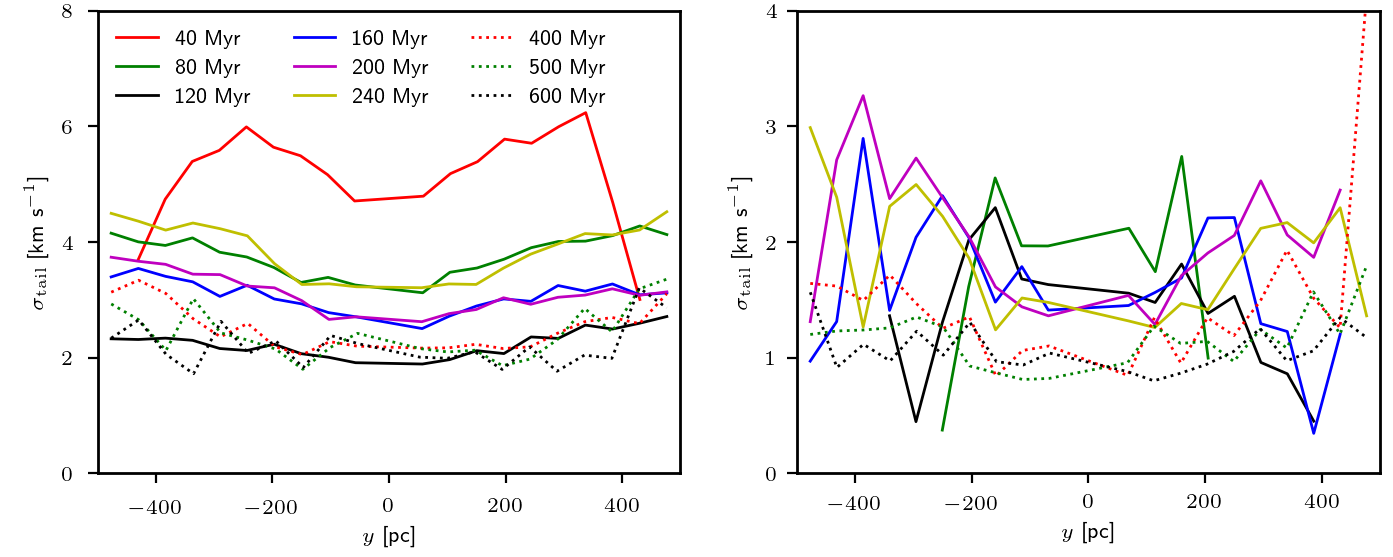}
\caption{Tail velocity dispersion $\sigma_{\rm tail}$ as a function of the distance $y$ along the tail 
for model C10G13 (left panel) and model C10W1 (right panel) at selected times as indicated by lines. 
At a given position $y$, the value of $\sigma_{\rm tail}$ evolves non-monotonically for model C10G13; 
it decreases first, reaching minima around $120 \Myr$, whereupon it starts increasing.
}
\label{fVelDispProf}
\end{figure*} \else \fi

\iffigs
\begin{figure*}[h!]
\includegraphics[width=\textwidth]{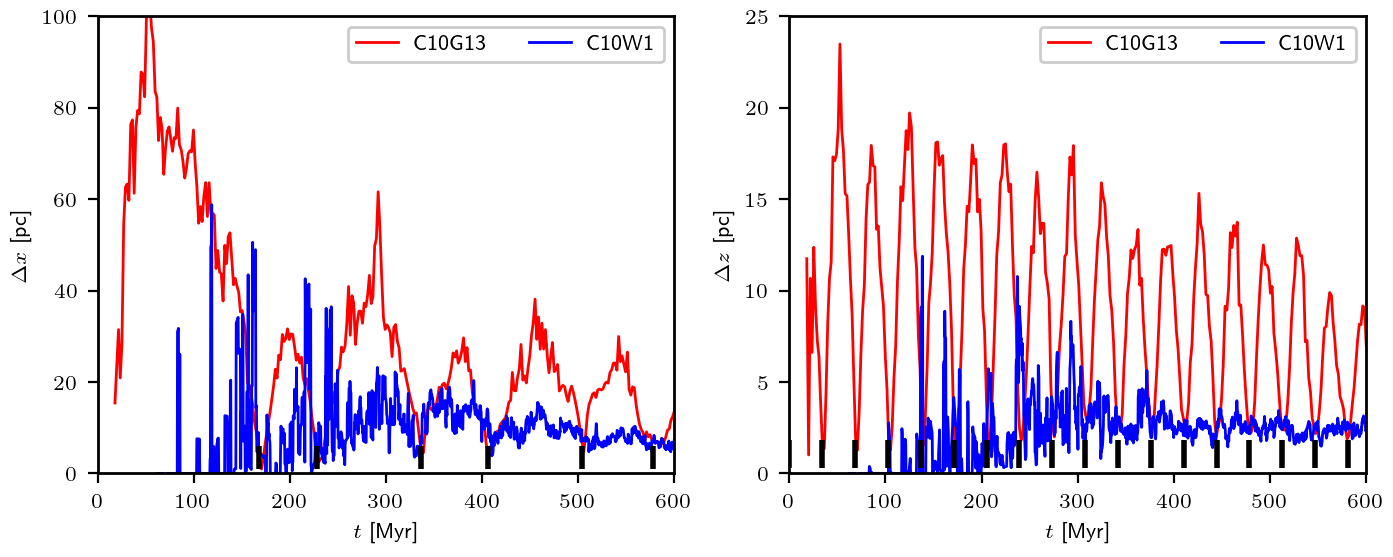}
\caption{Evolution of tail thickness $\Delta x$ (left panel) and tail thickness $\Delta z$ (right panel) 
for models C10G13 (red lines) and C10W1 (blue lines). 
The analysis was done at the distance $200 \Pc < y < 300\Pc$ from the cluster.
The predictions of the times of tail thickness minima according to eqs. (\ref{eTailThickness2}) and (\ref{eeOrbitZ}) 
are indicated by the black bars.}
\label{fthickness}
\end{figure*} \else \fi

\iffigs
\begin{figure*}[h!]
\includegraphics[width=\textwidth]{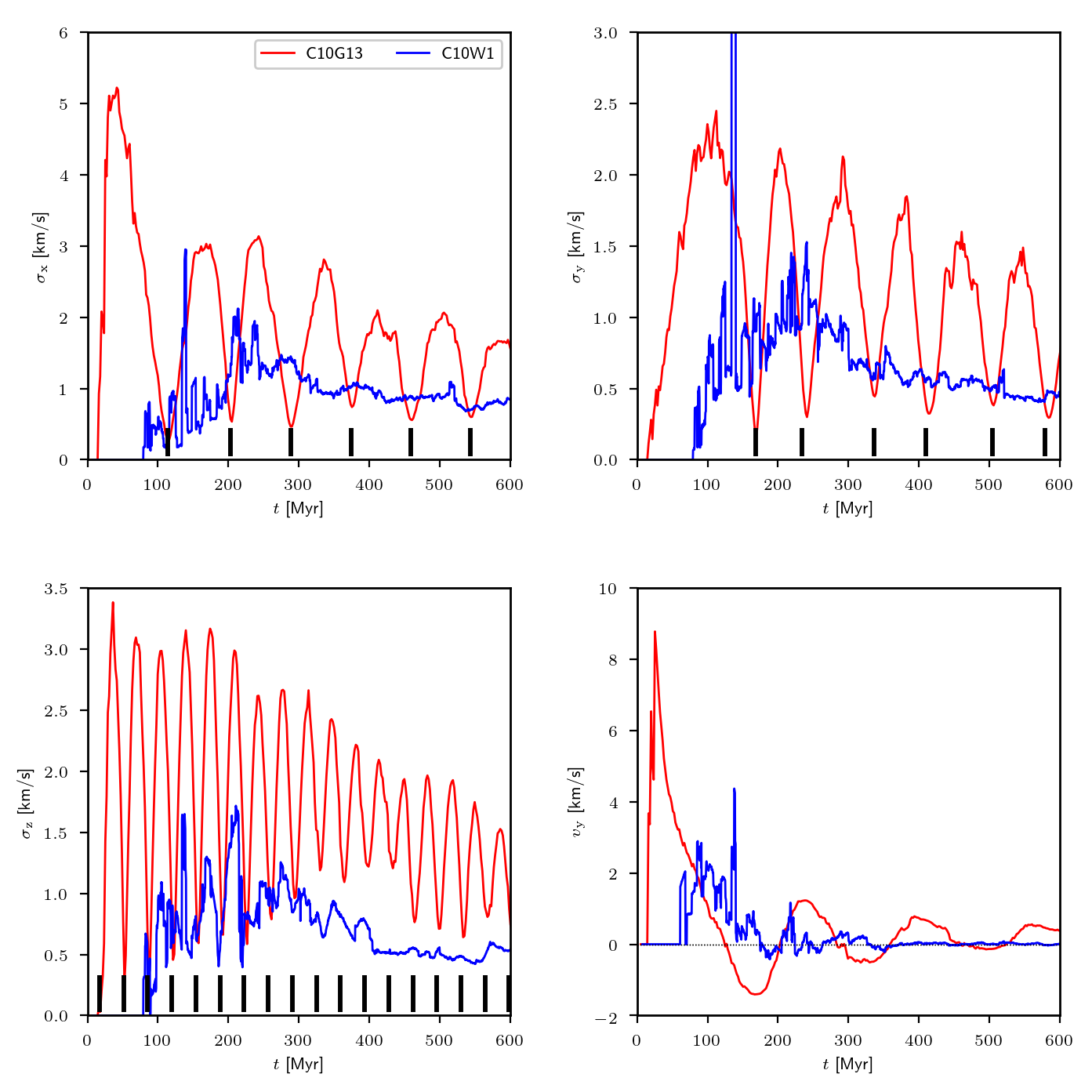}
\caption{Comparison of the tail velocity structure between models C10G13 (red lines) and C10W1 (blue lines) at $200 \Pc < y < 300\Pc$. 
\figpan{Upper left:} Velocity dispersion $\sigma_{\rm x}$.  
\figpan{Upper right:} Velocity dispersion $\sigma_{\rm y}$.  
\figpan{Lower left:} Velocity dispersion $\sigma_{\rm z}$.  
The minima of velocity dispersions, as calculated from eqs. (\ref{eTailSigmaY2}), (\ref{eTailSigmaX2}), 
and (\ref{eeOrbitZ}), are indicated by the black bars. 
\figpan{Lower right:} The $v_{\rm y}$ component of the mean bulk tail velocity. 
}
\label{fvel_vs_time}
\end{figure*} \else \fi

The relatively low escape speed of stars forming tail II results in the quadrant $x>0$ and $y>0$ and the quadrant $x<0$ and $y<0$ 
being practically devoid of stars (the very few stars occupying this area
 are mainly fast ejectors).
In contrast, tail I populates these quadrants substantially 
(cf. the upper middle panel of \reff{fvfield_2columns}) apart from the time near the minimum thickness.
Near the events of the minimum tail thickness $t_{\rm x0}$, tail I is almost straight, while 
tail II is clearly S-shaped at the star cluster, and both tails can be easily discerned (lower middle row 
of \reff{fvfield_2columns}).


%

\subsection{The kinematics of tail stars}

The right panels of Fig. \ref{fvfield_2columns} show the directions of motion of stars in the tail, where 
the colour shows the direction of motion measured from the positive $x$ axis clockwise as
indicated by the colourbar and the coloured arrows. 
Note that this figure does not provide the absolute value of the speed, which is discussed in \refs{ssBulkExp}. 
Near the beginning of the evolution, tail I expands radially (the upper row), and gets 
rotated clockwise due to the Coriolis force, gradually elongating the tail along the axis $x = 0$.  
Later (the upper and upper middle plot), tail I 
has a shear-like motion at the time between the minima of $x$ thickness, where
the stars located at $R > R_0$ ($R_0$ is the radius of the cluster orbit around the Galaxy, which corresponds to 
$x = 0$) move slower than the cluster, and stars at $R < R_0$ move faster than the cluster. 
The description of this movement as a shear is only an approximation, the trajectories are more complicated (cf. the arrows), 
where stars further away from the cluster tend to move radially (e.g. the stars in the upper panel at $y \approx -300 \Pc$). 
The velocity structure of tail I changes near times $t_{\rm x0}$, where the 
shear-like movement transforms to an outward motion. 

The majority of stars in the tail of the gas free model C10W1  
moves outwards from the cluster (lower plot); the few stars moving in the
opposite direction are near the cusps on their epicycles. 
Tail II of the gas rich model C10G13 (upper middle and lower middle rows) is of the similar velocity structure, 
demonstrating that tail II is kinematically simpler than tail I. 

To summarise, the trajectory of individual stars in either tail I or tail II is approximately 
an epicycle, but what causes the differences in the tail morphology is the value of the escape speed and 
the duration of the interval during which the stars escaped the cluster. 
The escape speed sets the semi-minor axes $X$ of the epicycle as well as the orbital radius of the guiding cetre $R_{\rm g}$ 
(eqs. \ref{eR1} and \ref{exv}), which determines the shape of the tail; if $X \lesssim r_{\rm t}$ the tail is S-shaped, 
if $X \gtrsim r_{\rm t}$ the tail is extended in the $x$ direction to several $r_{\rm t}$ without being S-shaped. 

\subsection{The extent of the tails}

The extent of the tidal tail is measured by 
finding a radius of a sphere centred on the cluster, which contains a given fraction of the total number of stars in the tail.
The stars bound to the cluster are not taken into account.
These radii are analogues to the concept of Lagrangian radii.
We prefer to calculate the radii for number instead of mass as we intend to 
avoid fluctuations due to a small number of massive stars.
The evolution of the radius containing $20$ \%, $50$ \% and $80$ \% stars is shown in Fig. \ref{flagrtail}.
Model C10G13 has a substantially longer tidal tail than model C10W1; measured as the tail 
half-number radius $r_{\rm h,tail}$, it is $r_{\rm h,tail} = 150 \Pc$ and $r_{\rm h,tail} = 550 \Pc$ 
for models C10W1 and C10G13, respectively at $t = 300 \Myr$.

The $r_{\rm h,tail}$ calculated semi-analytically according to \eq{erhsemiAnl} and scaled by 
the mean escape velocity $\widetilde{v}_{\rm e,I} = 4 \Kms$
is indicated by the dashed line. 
The semi analytical model follows the direct N-body simulations well, including a close approximation to  
the local maximum (at $\approx 120 \Myr$) and minimum (at $\approx 180 \Myr$) of the tail extent.
This implies that a good estimate for $r_{\rm h,tail}$ 
of any cluster (e.g. a substantially more massive cluster) at any time can be obtained by multiplying the 
semi-analytic result of \eq{erhsemiAnl} (as shown in the upper panel of \reff{frh_semiAnl}) by $\tilde{v}_{\rm e}/1 \Kms$, 
where $\tilde{v}_{\rm e}$ is the expected median velocity of escaping stars. 

Figure \ref{fDensProf} shows the stellar number density per $1 \Pc$ length of the tail as a function of the distance $y$ from the
cluster at selected time instants as indicated by lines.
Model C10G13 (left panel) swiftly builds a long tail.
After $\approx 80 \Myr$, the density in the tail decreases only slightly with increasing $|y|$
(the density drops only by a factor of $1.4$ between $y = 50 \Pc$ and $y = 200 \Pc$ at $t = 240 \Myr$),
so the tail has long wings.
At a given position $y$, the density decreases very slowly with time.
In contrast, model C10W1 (right panel) builds the tidal tail very slowly, and the tail density
sharply decreases outwards (the density drops by a factor of $4$ between $y = 50 \Pc$ and $y = 200 \Pc$ at $y = 240 \Myr$).

\subsection{The epicyclic overdensities}

Next, we search for epicyclic overdensities in the tidal tails.
\citet{Kupper2008} derive an analytic formula (their eqs. 5 and 7) for the distance $y_{\rm C}$ of the first epicyclic
overdensity from the star cluster; $y_{\rm C} = \pm 2 \pi \gamma (1 - \gamma^2) r_{\rm t}$.
The formula was derived under two simplified assumptions:(i) the cluster does not gravitationally influence a
star after it escapes; (ii) stars escape from the cluster with almost zero velocity.
The potential adopted in this work has $\gamma = 2\omega/ \kappa = 1.41$ (cf. \reff{fGalModel}), 
for which $y_{\rm C} = \pm 210 \Pc$.
This distance is indicated by the vertical dashed lines in Fig. \ref{fDensProf}.
Note that for this potential, $y_{\rm C} \approx \pm 9 r_{\rm t}$, is substantially smaller than
for the Keplerian potential with $\gamma = 2$ used by \citet{Kupper2008}, where $y_{\rm C} = \pm 12 \pi r_{\rm t}$.
Our models have no sign of epicyclic overdensities by the time $\approx 300 \Myr$.

Model C10W1 builds the first epicyclic overdensity at a time of $600 \Myr$.
Model C10G13 also builds the first epicyclic overdensity after $\approx 400 \Myr$, albeit slightly closer
to the cluster (around $180 \Pc$) than in model C10W1.
The positions of epicyclic overdensities, which are shown by the dashed lines, 
are in an excellent agreement with the analytic prediction of \citet{Kupper2008}.
The later formation of the overdensities than in the work of \citet{Kupper2008}, where the overdensities
form at $\approx 2 \times 2\pi/\kappa \approx 350 \Myr$,
might be caused by the difference of the IMF adopted in the models.
While we use a realistic IMF, \citet{Kupper2008} assume all stars to have the same mass of $1 \Msun$.
The presence of relatively massive stars in our models results in mass segregation with more
hard encounters between stars, which produces escapers with a broader velocity distribution and thus 
larger epicycle semi-minor axes $X$ \eqp{ex} 
as well as epicycles of larger loops, which smear the overdensities near the epicyclic cusps.

\subsection{The bulk expansion velocity and the velocity dispersion}

\label{ssBulkExp}

The mean value of the velocity $v_{\rm y}$ as a function of the $y$ distance from the cluster for the extreme models 
is shown in Fig. \ref{fvelalongx}.
Model C10G13 (left panel) has a very simple velocity structure of the tail. 
At a given time, the tail velocity linearly increases with cluster distance with proportionality constant $t_{\rm tail}$, 
i.e. $v_{\rm y} = y/t_{\rm tail}$. 
The constant $t_{\rm tail}$ varies with time, changing the slope and also the sign of $v_{\rm y}$. 
The stars first recede from the cluster, but after $\approx 120 \Myr$, they start temporarily approaching the cluster, 
and at $\approx 215 \Myr$, the stars again start receding from the cluster. 
These events correspond to tail contraction and stretching as seen previously in \reff{flagrtail}.
At given position $y$, the amplitude of oscillations $v_{\rm y}$ decreases with time. 
Note that at given $y$, the stars move in the shear as described in Fig. \ref{fvfield_2columns}, 
but their average velocity $v_{\rm y}$ is so balanced that $v_{\rm y}$ always depends linearly on $y$.

The dependence $v_{\rm y} \propto y$ closely follows the properties of \eq{eTailSigmaX}, 
which is derived analytically, but which holds only at the specific times $t = 2 \pi l/ \kappa$. 
We plot $v_{\rm y}$ according to \eq{eTailSigmaX} for $t = 168 \Myr$, i.e. for $l = 1$ in the left panel of Fig. \ref{fvelalongx}, 
which is shown by the dashed line.
The analytical solution, which holds at $t = 168 \Myr$, is very close to the results of the simulation which is 
shown at $t = 160 \Myr$.

The tail velocity of the gas-free model C10W1 (right panel of Fig. \ref{fvelalongx}) shows a clear  
dependence of $v_{\rm y} \propto y$ only near the cluster and at the beginning of the evolution (up to $\approx 120 \Myr$). 
After that time, the velocity structure indicates a more complicated dependence on $y$. 
The epicyclic overdensity, which develops at $\pm 240 \Pc$ after $\approx 400 \Myr$, can be seen as the place where 
the outward motion of the cluster slows or even reverses to an inward motion. 
Also note that the value of $v_{\rm y}$ is at least by a factor of $3$ smaller than that of the cluster C10G13. 
Thus, a tail where the velocity increases as $v_{\rm y} \propto y/t_{\rm tail}$ to a larger distance from the cluster 
will point to a model strongly affected by gas expulsion. 
Note that the constant $t_{\rm tail}$ is provided by the simulations of gas-rich clusters, and 
its value for clusters of different $\tilde{v}_{\rm e,I}$ could be obtained only by a scaling of present models.



The tail velocity dispersion $\sigma_{\rm tail} = \sqrt{\sigma_{\rm x}^2 + \sigma_{\rm y}^2 + \sigma_{\rm z}^2}$ as a 
function of $y$ is shown in \reff{fVelDispProf}. 
The value of $\sigma_{\rm tail}$ evolves non-monotonically with time for model C10G13 (left panel); 
at given $y$, $\sigma_{\rm tail}$ decreases first, 
reaching its minimum at $\approx 120 \Myr$, and increasing afterwards. 
At any time, $\sigma_{\rm tail}$ shows only mild variations along $|y|$ with a slight increase with $|y|$. 
In contrast, model C10W1 (right panel) does not show the rapid time variability in $\sigma_{\rm tail}$.
Note that the value of $\sigma_{\rm tail}$ is by a factor of $\approx 2$ larger in model C10G13 than in C10W1.

\subsection{The minima of tail thickness and the minima of the $x-$, $y-$ and $z-$ velocity dispersion component}

After describing the characteristic properties of tail I and contrasting them with that of 
tail II, we turn our attention to the quantities which are intended for more direct comparison 
with the analytical model; namely, the minima of tail thickness and velocity dispersions. 
These quantities are measured at the distance $200 \Pc < y < 300 \Pc$ from the star cluster. 
Model C10G13 is a good representation of tail I as these stars dominate the number of tail stars. 
Model C10W1 represents tail II as this model does not form tail I at all.

The thickness of the tail is plotted in \reff{fthickness}; the left panel 
shows the thickness $\Delta x$ in the direction $x$, the right panel the thickness $\Delta z$ in the direction $z$. 
The thickness is an analogue to the half number radius defined above; 
it is defined as the distance from the tail centre which contains 50\% of the tail stars. 
The tail of model C10G13 (red line) shows rapid oscillations, and the times of the minima of these oscillations 
are in excellent agreement with the predicted minima of the analytical 
formulae, \eq{eTailThickness2} and \eq{eeOrbitZ} (thick black bars). 
Also note that while the minima of $\Delta z$ occur periodically, 
the minima of $\Delta x$ do not occur in periodic intervals. 
In contrast the tail thickness of model C10W1 (blue line) does not oscillate periodically; 
apart from several spikes caused by fast ejectors, its thickness fluctuates around a constant. 

The velocity structure of the tail is shown in \reff{fvel_vs_time}. 
As predicted by the (semi-)analytic model, the amplitude of the velocity dispersion for tail I oscillates with time 
(red line on the first three plots). 
The minima of the velocity dispersions are again in excellent agreement with the analytic predictions 
(eqs. \ref{eTailSigmaY2}, \ref{eTailSigmaX2}, and \ref{eeOrbitZ}).  
Tail II (blue line) shows no oscillations in velocity dispersions.  

The lower right panel of \reff{fvel_vs_time} shows the bulk velocity $v_{\rm y}$ of the tail. 
For model C10G13, the bulk velocity oscillates from positive values of higher amplitudes to 
negative values of lower amplitudes. 
It means that the overall tail expansion changes to tail contraction (the first event of tail 
contraction occurs between $120 \Myr$ and  $180 \Myr$). 
We already saw the contractions in \reff{flagrtail}. 
In contrast, tail II of model C10W1 shows almost no oscillations. 
The value of $|v_{\rm y}|$ is also substantially lower for model C10W1 ($0.1 \Kms$) than 
for model C10G13 ($1 \Kms$). 

\subsection{The validity of the approximation for clusters with lower velocity dispersion}

\label{ssSmallerSigma}

\iffigs
\begin{figure} 
\includegraphics[width=\columnwidth]{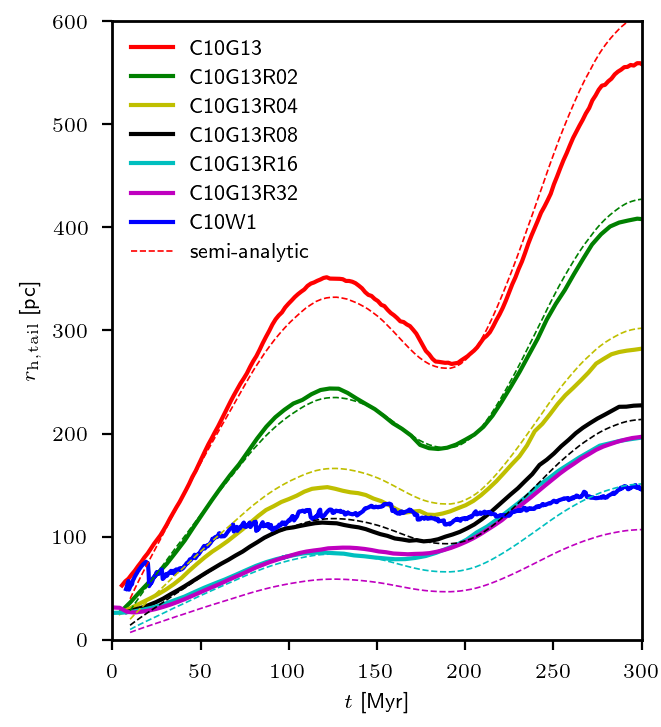}
\caption{
Evolution of half-number radius of the tail for models of different values of the initial velocity dispersion (cf. \reft{tVelApprox}). 
The semi-analytical solution \eqp{erhsemiAnl} is plotted by the dashed lines.
The result of eq. (\ref{erhsemiAnl}) is scaled to $v_{\rm e} = 4 \Kms$ for model C10G13, and then decreased by a factor of 
$N^{0.5}$ for models labelled C10G13R"N" to scale with the initial velocity dispersion of the cluster. 
}
\label{flagrtail_lvd}
\end{figure} \else \fi

\iffigs
\begin{figure*}[h!]
\includegraphics[width=\textwidth]{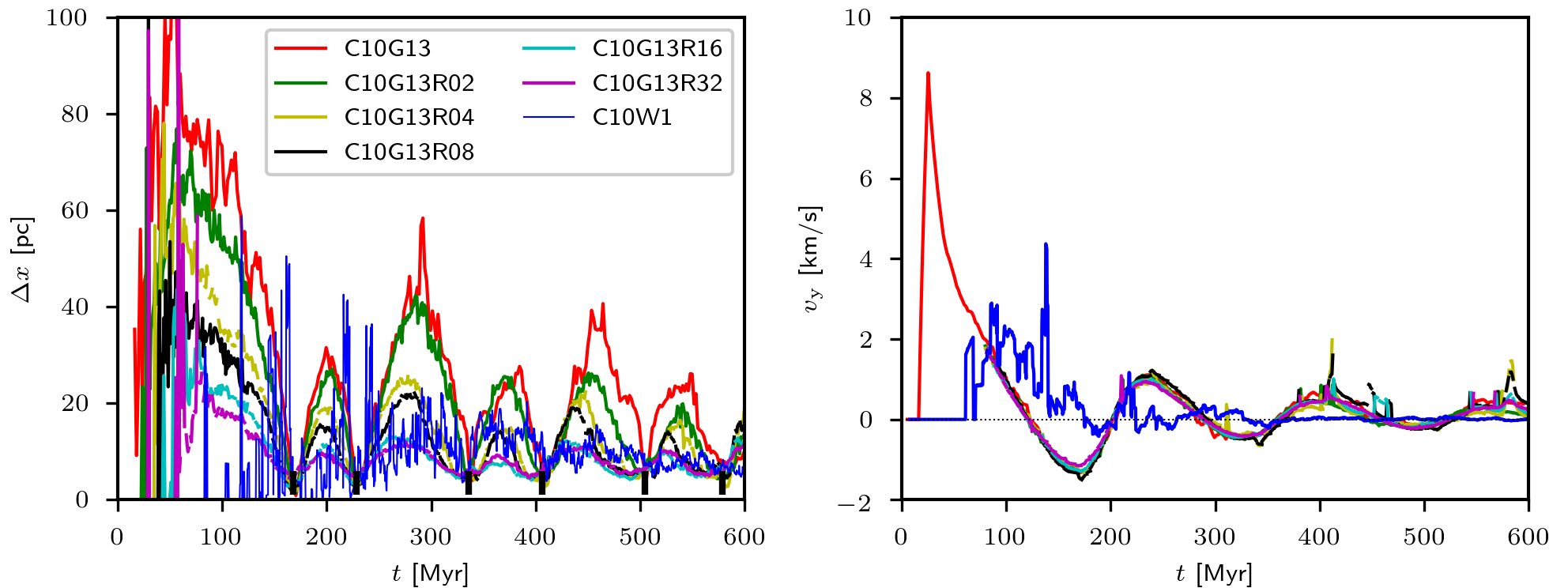}
\caption{
\figpan{Left panel:} Evolution of tail thickness $\Delta x$ for models of different values of the initial velocity dispersion. 
The events of the minimum of tail thickness \eqp{eTailThickness2} are indicated by the black bars. 
\figpan{Right panel:} Evolution of the bulk velocity $v_{\rm y}$. 
For clarity, we do not plot the first $80 \Myr$ of evolution for some of the models.  
}
\label{ftthick_bulk_vely}
\end{figure*} \else \fi

\begin{table}
\begin{tabular}{ccc}
Model name & $a_{\rm cl}$ [$\Pc$] & $\sigma_{\rm cl}$ [$\Kms$] \\
\hline
C10G13 & 0.23 & 8.5 \\
C10G13R02 & 0.46 & 6.0 \\
C10G13R04 & 0.92 & 4.3 \\
C10G13R08 & 1.84 & 3.0 \\
C10G13R16 & 3.68 & 2.1 \\
C10G13R32 & 7.36 & 1.5
\end{tabular}
\caption{
Parameters of the simulations described in \refs{ssSmallerSigma}. 
The Plummer length is coded by the model name, for example model C10G13R"N", has the parameter $a_{\rm cl}$ "N" times larger than that of the control model C10G13.
The velocity $\widetilde{v}_{\rm e,II}$ is $2.2 \Kms$ for all the clusters. 
}
\label{tVelApprox}
\end{table}

One of the assumptions in finding the (semi-)analytic formulae for the tail I is that the velocity dispersion, $\sigma_{\rm cl}$, of the cluster prior to gas expulsion 
is significantly larger than the velocity, $\widetilde{v}_{\rm e,II}$, corresponding to the potential difference at the surface of the Jacobi sphere. 
Here, we test the validity of the formulae in a non-ideal case when the assumption $\sigma_{\rm cl} \gg \widetilde{v}_{\rm e,II}$ is relaxed. 

For the purpose of this task, we construct additional cluster models, which are identical to model C10G13, with the exception that 
they have larger Plummer lengths $a_{\rm cl}$ and $a_{\rm gas}$, and a longer gas expulsion time-scale $\tau_{\rm M}$, 
which is given by $\tau_{\rm M} = a_{\rm gas}/(10 \Kms)$. 
The time delay $t_{\rm d}$ is again $0.6 \Myr$. 
These models are listed in \reft{tVelApprox}. 

\reff{flagrtail_lvd} contrasts the half-number radius of the tidal tail of the simulated clusters (solid lines) 
with the approximation of \eq{erhsemiAnl} (dashed lines). 
The models with $a_{\rm cl} \lesssim 1.84 \Pc$ ($\sigma_{\rm cl} \gtrsim 3 \Kms$) follow closely the semi-analytic solution. 
Models with $a_{\rm cl} \gtrsim 3.68 \Pc$ ($\sigma_{\rm cl} \lesssim 2.1 \Kms$) differ from the semi-analytic solution, but never more than by a factor of 2. 
Equation (\ref{erhsemiAnl}) is thus valid for $\sigma_{\rm cl} \gtrsim \widetilde{v}_{\rm e,II}$ ($\widetilde{v}_{\rm e,II}$ is $2.2 \Kms$). 
Note that the model without gas expulsion (C10W1), which has $a_{\rm cl}$ smaller than models C10G13R08 to C10G13R32, 
does not show the decrease in $r_{\rm h,tail}$ between $120 \Myr$ and $180 \Myr$ followed by an increase, which is pronounced in the models with gas expulsion; 
the characteristic shape of the models with gas expulsion is caused by the process of gas expulsion, not by the large initial radius of these clusters. 

From the other major tail quantities shown in Figs. \ref{fDensProf} to \ref{fvel_vs_time}, we depict the 
$\Delta x$ tail thickness and the bulk velocity $v_{\rm y}$ for brevity (\reff{ftthick_bulk_vely}); the other quantities show qualitatively similar behaviour. 
All the models with gas expulsion show aperiodic oscillations in $\Delta x$ (left panel of \reff{ftthick_bulk_vely}) where the amplitude 
of oscillation decreases with increasing $a_{\rm cl}$. 
The minima of tail x- thickness occur close to those predicted by \eq{eTailThickness2} for all the gas expulsion models, even the ones 
with $\sigma_{\rm cl} \lesssim \widetilde{v}_{\rm e,II}$. 
The bulk tail velocity $v_{\rm y}$ has its characteristic damped oscillations for all the models with gas expulsion (right panel of \reff{ftthick_bulk_vely}),  
distinct from the evolution of $v_{\rm y}$ for the model C10W1.
These results indicate that all the semi-analytic formulae present in \refs{stailAnal} are valid for any 
$\sigma_{\rm cl} \gtrsim \widetilde{v}_{\rm e,II}$.

\section{Conclusions}

\label{sSummary}

We study the morphology and kinematics of a tidal tail (tail I) formed due to a rapid release of stars 
from a star cluster, which orbits in a realistic gravitational potential of the Galaxy.
The mechanism responsible for releasing the stars can be, for example, gas expulsion out of an embedded star 
cluster by stellar feedback. 
Tail I has distinct kinematical and morphological signatures from the more often studied tidal tails formed due to gradual 
dynamical evaporation or ejections (tail II), which occur throughout the cluster life-time. 
These signatures are studied by a (semi-)analytic model and direct simulations with the \nbdvi code. 

The main differences between tail I and tail II are as follows:
\begin{itemize}
\item
The thickness of tail I oscillates aperiodically with time in direction $x$ and periodically in direction $z$ 
(cf. \reff{fConfig} for orientation of the axes).
The thickness varies substantially from several $\Pc$ at the minima to $\approx 30\Pc$ at the maxima.
In contrast, the $x$ or $z$ thickness of tail II does not show these minima (cf. \reff{fthickness}). 
\item
At the times of the minimum thickness, the narrow tail I is either parallel to the axis $y$, 
or it is tilted by an angle $\lesssim 20 ^{\circ}$ to the $y$ axis. 
When the tail is tilted, its leading part always points towards the inner Galaxy.
As the tail evolves, the two types of minima thickness alternate in time with the former case (tail parallel to the axis $y$) 
occurring for an odd minimum followed by the latter case (tail tilted to the axis $y$) occurring for an even minimum. 
\item
The velocity dispersions in directions $x$, $y$, $z$ also show rapid changes with time for tail I, but these 
quantities are non-oscillating for tail II. 
\item
The time events of the thickness minima, the angle of the tail relative to the axis $y$, 
and velocity dispersion minima of tail I can be found analytically, 
and they very closely agree with the corresponding \nbdvi calculations. 
These quantities also depend only on the three galactic frequencies $\omega$, $\kappa$ and $\nu$,
being unrelated to any property of the cluster. 
\item
The Lagrange radii of tail I behave non-monotonically, with longer epochs of expansion interspersed with shorter 
epochs of contraction. 
A good approximation to the evolution of the half-mass tail radius $r_{\rm h,tail}$ can be found semi-analytically. 
The value of $r_{\rm h,tail}$ at a given time is only a function of the stellar escape speed, $v_{\rm e}$, and Galactic 
frequencies $\omega$, $\kappa$ and $\nu$. 
\item
The stellar density of tail I shows only a mild decrease with $|y|$, so the tail has long wings (cf. \reff{fDensProf}). 
In contrast, the stellar density of tail II decreases sharply with $|y|$. 
\item
The velocity component $v_y$ of tail I increases linearly with the distance from the cluster, 
i.e. $v_y =  y/t_{\rm tail}$ (cf. \reff{fvelalongx}), where the proportionality constant $t_{\rm tail}$ evolves 
(and increases in absolute value on average) with time. 
In contrast, tail II has a more complicated velocity structure, particularly at later times ($t \gtrsim 400 \Myr$), when 
epicycle overdensities develop (right panel of \reff{fvelalongx}).
\end{itemize}

The semi-analytical model indicates that the different morphology and kinematics of tail I 
as contrasted to that of tail II is due to the two following reasons: 
(i) The stars of tail I leave the cluster during a short interval in 
comparison to the epicyclic time $2 \pi /\kappa$. 
This means that all the tail I stars have nearly the same value of $\kappa t$ in \eq{eOrbit}. 
This fact then simplifies the structure of the ensuing equations, which enables the tail $x$ thickness 
and the $x$, $y$ and $z$ component of the velocity dispersion to reach zero at distinct times. 
In contrast, tail II forms gradually with nearly a constant rate over several $2 \pi /\kappa$, so its stars 
have different factors $\kappa t$ in \eq{eOrbit}.
(ii) The stars of tail I leave the cluster with escape velocities substantially larger than the velocity corresponding to the potential difference 
around the sphere of the Jacobi radius. 
This means that these stars escape the cluster almost isotropically, and that the sizes of their epicycle semi-minor axes 
are larger than the Jacobi radius.
These stars thus form a tail of thickness of several Jacobi radii in the direction $x$, basically enveloping 
tail II (cf. upper middle row of \reff{fvfield_2columns}). 

In contrast, stars in tail II leave the cluster with escape velocities usually lower than the potential difference 
around the sphere of the Jacobi radius. 
This means that these stars can escape the cluster only in the vicinity of Lagrange points L1 or L2, and that 
the sizes of the epicycle semi-minor axes are smaller or comparable to that of the Jacobi radius. 
These stars form relatively thin, slowly expanding and S-shaped tail II. 
Numerical experiments demonstrate that all the (semi-)analytic estimates stated 
in the paper are valid under weaker conditions than given by (ii); it is sufficient that 
the velocities $\sigma_{\rm cl}$ and $\widetilde{v}_{\rm e,II}$ are comparable. 
This implies that the estimates hold for a larger set of initial conditions, particularly for initially less concentrated clusters. 

The closer agreement between the semi-analytic model for tail I and the direct numerical simulation 
enables to estimate the basic shape of the possible tail I for any Galactic star cluster  
on a circular orbit in the Galactic plane. 
For a cluster of age $t$ on a circular orbit with given frequencies $\omega$, $\kappa$ and assuming escape 
speed $v_{\rm e}$, 
the half-mass radius of the tail can be estimated from \eq{erhsemiAnl} (or upper panel of \reff{frh_semiAnl}). 
These estimates are calculated for a cluster orbiting on the Solar galactocentric radius $8.5 \Kpc$ 
as well as at radii $6.5 \Kpc$ and $10.5 \Kpc$ and in the plane ($z=0$) of the Galactic disc. 
The approximative thickness of the tail could be obtained by comparing the cluster age 
$t$ with the events of tail minima thickness for the cluster orbit from \eq{eTailThickness2}.
An observational detection of tail I would strongly favour early rapid gas expulsion from the cluster 
with a relatively low SFE.

The present numerical study is extended by more parameters of the conditions of the early gas expulsion in Paper II, 
where we also provide direct predictions for the structure of the tidal tails of the Pleiades star cluster 
for different scenarios for the early gas expulsion.

\begin{acknowledgements}

The authors express thanks to Sverre Aarseth for his development of the code \nbdvid, 
which was vital for the present simulations as well as for his useful comments on the manuscript. 
FD would like to thank Stefanie Walch for her encouragement to work on the project while being employed within her group.
FD acknowledges the support by the Collaborative Research Centre 956, sub-project C5, funded by the Deutsche Forschungsgemeinschaft (DFG) – project ID 184018867.

\end{acknowledgements}

%
%

\bibliographystyle{aa} 
\bibliography{clusterTail} 

\begin{thebibliography}{69}
\expandafter\ifx\csname natexlab\endcsname\relax\def\natexlab#1{#1}\fi

\bibitem[{{Aarseth}(1999)}]{Aarseth1999}
{Aarseth}, S.~J. 1999, \pasp, 111, 1333

\bibitem[{{Aarseth}(2003)}]{Aarseth2003}
{Aarseth}, S.~J. 2003, {Gravitational N-Body Simulations} (Cambridge: Cambridge
  University Press)

\bibitem[{{Aarseth} {et~al.}(1974){Aarseth}, {Henon}, \&
  {Wielen}}]{Aarseth1974b}
{Aarseth}, S.~J., {Henon}, M., \& {Wielen}, R. 1974, \aap, 37, 183

\bibitem[{{Aarseth} \& {Zare}(1974)}]{Aarseth1974a}
{Aarseth}, S.~J. \& {Zare}, K. 1974, Celestial Mechanics, 10, 185

\bibitem[{{Ahmad} \& {Cohen}(1973)}]{Ahmad1973}
{Ahmad}, A. \& {Cohen}, L. 1973, Journal of Computational Physics, 12, 389

\bibitem[{{Allen} \& {Santillan}(1991)}]{Allen1991}
{Allen}, C. \& {Santillan}, A. 1991, \rmxaa, 22, 255

\bibitem[{{Alves} {et~al.}(2007){Alves}, {Lombardi}, \& {Lada}}]{Alves2007}
{Alves}, J., {Lombardi}, M., \& {Lada}, C.~J. 2007, \aap, 462, L17

\bibitem[{{Arnold}(1989)}]{Arnold1989}
{Arnold}, V.~I. 1989, {Mathematical Methods of Classical Mechanics (2nd ed.;
  New York: Springer)}

\bibitem[{{Banerjee} \& {Kroupa}(2013)}]{Banerjee2013}
{Banerjee}, S. \& {Kroupa}, P. 2013, \apj, 764, 29

\bibitem[{{Banerjee} \& {Kroupa}(2017)}]{Banerjee2017}
{Banerjee}, S. \& {Kroupa}, P. 2017, \aap, 597, A28

\bibitem[{{Bate}(2012)}]{Bate2012}
{Bate}, M.~R. 2012, \mnras, 419, 3115

\bibitem[{{Bate}(2014)}]{Bate2014}
{Bate}, M.~R. 2014, \mnras, 442, 285

\bibitem[{{Baumgardt} {et~al.}(2008){Baumgardt}, {De Marchi}, \&
  {Kroupa}}]{Baumgardt2008}
{Baumgardt}, H., {De Marchi}, G., \& {Kroupa}, P. 2008, \apj, 685, 247

\bibitem[{{Baumgardt} \& {Kroupa}(2007)}]{Baumgardt2007}
{Baumgardt}, H. \& {Kroupa}, P. 2007, \mnras, 380, 1589

\bibitem[{{Baumgardt} \& {Makino}(2003)}]{Baumgardt2003}
{Baumgardt}, H. \& {Makino}, J. 2003, \mnras, 340, 227

\bibitem[{{Binney} \& {Tremaine}(2008)}]{Binney2008}
{Binney}, J. \& {Tremaine}, S. 2008, {Galactic Dynamics: Second Edition}
  (Princeton University Press)

\bibitem[{{Boily} \& {Kroupa}(2003{\natexlab{a}})}]{Boily2003a}
{Boily}, C.~M. \& {Kroupa}, P. 2003{\natexlab{a}}, \mnras, 338, 665

\bibitem[{{Boily} \& {Kroupa}(2003{\natexlab{b}})}]{Boily2003b}
{Boily}, C.~M. \& {Kroupa}, P. 2003{\natexlab{b}}, \mnras, 338, 673

\bibitem[{{Churchwell}(2002)}]{Churchwell2002}
{Churchwell}, E. 2002, \araa, 40, 27

\bibitem[{{Dale} \& {Bonnell}(2011)}]{Dale2011}
{Dale}, J.~E. \& {Bonnell}, I. 2011, \mnras, 414, 321

\bibitem[{{Dale} {et~al.}(2012){Dale}, {Ercolano}, \& {Bonnell}}]{Dale2012}
{Dale}, J.~E., {Ercolano}, B., \& {Bonnell}, I.~A. 2012, \mnras, 424, 377

\bibitem[{{De Marchi} {et~al.}(2007){De Marchi}, {Paresce}, \&
  {Pulone}}]{DeMarchi2007}
{De Marchi}, G., {Paresce}, F., \& {Pulone}, L. 2007, \apjl, 656, L65

\bibitem[{{Fujii}(2015)}]{Fujii2015b}
{Fujii}, M.~S. 2015, \pasj, 67, 59

\bibitem[{{Fujii} \& {Portegies Zwart}(2011)}]{Fujii2011}
{Fujii}, M.~S. \& {Portegies Zwart}, S. 2011, Science, 334, 1380

\bibitem[{{Fujii} \& {Portegies Zwart}(2015)}]{Fujii2015a}
{Fujii}, M.~S. \& {Portegies Zwart}, S. 2015, \mnras, 449, 726

\bibitem[{{Gavagnin} {et~al.}(2017){Gavagnin}, {Bleuler}, {Rosdahl}, \&
  {Teyssier}}]{Gavagnin2017}
{Gavagnin}, E., {Bleuler}, A., {Rosdahl}, J., \& {Teyssier}, R. 2017, \mnras,
  472, 4155

\bibitem[{{Geyer} \& {Burkert}(2001)}]{Geyer2001}
{Geyer}, M.~P. \& {Burkert}, A. 2001, \mnras, 323, 988

\bibitem[{{Goodwin}(1997)}]{Goodwin1997}
{Goodwin}, S.~P. 1997, \mnras, 284, 785

\bibitem[{{Goodwin} {et~al.}(2008){Goodwin}, {Nutter}, {Kroupa},
  {Ward-Thompson}, \& {Whitworth}}]{Goodwin2008}
{Goodwin}, S.~P., {Nutter}, D., {Kroupa}, P., {Ward-Thompson}, D., \&
  {Whitworth}, A.~P. 2008, \aap, 477, 823

\bibitem[{{Grudi{\'c}} {et~al.}(2018){Grudi{\'c}}, {Hopkins},
  {Faucher-Gigu{\`e}re}, {Quataert}, {Murray}, \& {Kere{\v s}}}]{Grudic2018}
{Grudi{\'c}}, M.~Y., {Hopkins}, P.~F., {Faucher-Gigu{\`e}re}, C.-A., {et~al.}
  2018, \mnras, 475, 3511

\bibitem[{{Haid} {et~al.}(2019){Haid}, {Walch}, {Seifried}, {W{\"u}nsch},
  {Dinnbier}, \& {Naab}}]{Haid2019}
{Haid}, S., {Walch}, S., {Seifried}, D., {et~al.} 2019, \mnras, 482, 4062

\bibitem[{{H{\'e}nault-Brunet}
  {et~al.}(2012{\natexlab{a}}){H{\'e}nault-Brunet}, {Evans}, {Sana}, {Gieles},
  {Bastian}, {Ma{\'{\i}}z Apell{\'a}niz}, {Markova}, {Taylor}, {Bressert},
  {Crowther}, \& {van Loon}}]{Henault2012}
{H{\'e}nault-Brunet}, V., {Evans}, C.~J., {Sana}, H., {et~al.}
  2012{\natexlab{a}}, \aap, 546, A73

\bibitem[{{H{\'e}nault-Brunet}
  {et~al.}(2012{\natexlab{b}}){H{\'e}nault-Brunet}, {Evans}, {Sana}, {Gieles},
  {Bastian}, {Ma{\'\i}z Apell{\'a}niz}, {Markova}, {Taylor}, {Bressert},
  {Crowther}, \& {van Loon}}]{Henault-Brunet2012}
{H{\'e}nault-Brunet}, V., {Evans}, C.~J., {Sana}, H., {et~al.}
  2012{\natexlab{b}}, \aap, 546, A73

\bibitem[{{Hills}(1980)}]{Hills1980}
{Hills}, J.~G. 1980, \apj, 235, 986

\bibitem[{{Hurley} {et~al.}(2000){Hurley}, {Pols}, \& {Tout}}]{Hurley2000}
{Hurley}, J.~R., {Pols}, O.~R., \& {Tout}, C.~A. 2000, \mnras, 315, 543

\bibitem[{{Kroupa}(2001)}]{Kroupa2001a}
{Kroupa}, P. 2001, \mnras, 322, 231

\bibitem[{{Kroupa} {et~al.}(2001){Kroupa}, {Aarseth}, \&
  {Hurley}}]{Kroupa2001b}
{Kroupa}, P., {Aarseth}, S., \& {Hurley}, J. 2001, \mnras, 321, 699

\bibitem[{{Kroupa} \& {Boily}(2002)}]{Kroupa2002}
{Kroupa}, P. \& {Boily}, C.~M. 2002, \mnras, 336, 1188

\bibitem[{{Kroupa} {et~al.}(2013){Kroupa}, {Weidner}, {Pflamm-Altenburg},
  {Thies}, {Dabringhausen}, {Marks}, \& {Maschberger}}]{Kroupa2013}
{Kroupa}, P., {Weidner}, C., {Pflamm-Altenburg}, J., {et~al.} 2013, {The
  Stellar and Sub-Stellar Initial Mass Function of Simple and Composite
  Populations}, ed. T.~D. {Oswalt} \& G.~{Gilmore}, p. 115

\bibitem[{{Kuhn} {et~al.}(2014){Kuhn}, {Feigelson}, {Getman}, {Baddeley},
  {Broos}, {Sills}, {Bate}, {Povich}, {Luhman}, {Busk}, {Naylor}, \&
  {King}}]{Kuhn2014}
{Kuhn}, M.~A., {Feigelson}, E.~D., {Getman}, K.~V., {et~al.} 2014, \apj, 787,
  107

\bibitem[{{Kuhn} {et~al.}(2015){Kuhn}, {Feigelson}, {Getman}, {Sills}, {Bate},
  \& {Borissova}}]{Kuhn2015}
{Kuhn}, M.~A., {Feigelson}, E.~D., {Getman}, K.~V., {et~al.} 2015, \apj, 812,
  131

\bibitem[{{Kuhn} {et~al.}(2019){Kuhn}, {Hillenbrand}, {Sills}, {Feigelson}, \&
  {Getman}}]{Kuhn2019}
{Kuhn}, M.~A., {Hillenbrand}, L.~A., {Sills}, A., {Feigelson}, E.~D., \&
  {Getman}, K.~V. 2019, \apj, 870, 32

\bibitem[{{K{\"u}pper} {et~al.}(2010){K{\"u}pper}, {Kroupa}, {Baumgardt}, \&
  {Heggie}}]{Kupper2010}
{K{\"u}pper}, A.~H.~W., {Kroupa}, P., {Baumgardt}, H., \& {Heggie}, D.~C. 2010,
  \mnras, 401, 105

\bibitem[{{K{\"u}pper} {et~al.}(2008){K{\"u}pper}, {MacLeod}, \&
  {Heggie}}]{Kupper2008}
{K{\"u}pper}, A.~H.~W., {MacLeod}, A., \& {Heggie}, D.~C. 2008, \mnras, 387,
  1248

\bibitem[{{Kustaanheimo} \& {Stiefel}(1965)}]{Kustaanheimo1965}
{Kustaanheimo}, P. \& {Stiefel}, E. 1965, Reine Angew. Math., 218, 204

\bibitem[{{Lada} \& {Lada}(2003)}]{Lada2003}
{Lada}, C.~J. \& {Lada}, E.~A. 2003, \araa, 41, 57

\bibitem[{{Lada} {et~al.}(1984){Lada}, {Margulis}, \& {Dearborn}}]{Lada1984}
{Lada}, C.~J., {Margulis}, M., \& {Dearborn}, D. 1984, \apj, 285, 141

\bibitem[{{Longmore} {et~al.}(2014){Longmore}, {Kruijssen}, {Bastian}, {Bally},
  {Rathborne}, {Testi}, {Stolte}, {Dale}, {Bressert}, \&
  {Alves}}]{Longmore2014}
{Longmore}, S.~N., {Kruijssen}, J.~M.~D., {Bastian}, N., {et~al.} 2014, in
  Protostars and Planets VI, ed. H.~{Beuther}, R.~S. {Klessen}, C.~P.
  {Dullemond}, \& T.~{Henning}, 291

\bibitem[{{L{\"u}ghausen} {et~al.}(2015){L{\"u}ghausen}, {Famaey}, \&
  {Kroupa}}]{Lughausen2015}
{L{\"u}ghausen}, F., {Famaey}, B., \& {Kroupa}, P. 2015, Canadian Journal of
  Physics, 93, 232

\bibitem[{{Makino}(1991)}]{Makino1991}
{Makino}, J. 1991, \apj, 369, 200

\bibitem[{{Marks} \& {Kroupa}(2012)}]{Marks2012}
{Marks}, M. \& {Kroupa}, P. 2012, \aap, 543, A8

\bibitem[{{Marks} {et~al.}(2008){Marks}, {Kroupa}, \& {Baumgardt}}]{Marks2008}
{Marks}, M., {Kroupa}, P., \& {Baumgardt}, H. 2008, \mnras, 386, 2047

\bibitem[{{Mathieu}(1983)}]{Mathieu1983}
{Mathieu}, R.~D. 1983, \apjl, 267, L97

\bibitem[{{Megeath} {et~al.}(2016){Megeath}, {Gutermuth}, {Muzerolle},
  {Kryukova}, {Hora}, {Allen}, {Flaherty}, {Hartmann}, {Myers}, {Pipher},
  {Stauffer}, {Young}, \& {Fazio}}]{Megeath2016}
{Megeath}, S.~T., {Gutermuth}, R., {Muzerolle}, J., {et~al.} 2016, \aj, 151, 5

\bibitem[{{Mikkola} \& {Aarseth}(1990)}]{Mikkola1990}
{Mikkola}, S. \& {Aarseth}, S.~J. 1990, Celestial Mechanics and Dynamical
  Astronomy, 47, 375

\bibitem[{{Miyamoto} \& {Nagai}(1975)}]{Miyamoto1975}
{Miyamoto}, M. \& {Nagai}, R. 1975, \pasj, 27, 533

\bibitem[{{Nutter} \& {Ward-Thompson}(2007)}]{Nutter2007}
{Nutter}, D. \& {Ward-Thompson}, D. 2007, \mnras, 374, 1413

\bibitem[{{Oh} {et~al.}(2015){Oh}, {Kroupa}, \& {Pflamm-Altenburg}}]{Oh2015}
{Oh}, S., {Kroupa}, P., \& {Pflamm-Altenburg}, J. 2015, \apj, 805, 92

\bibitem[{{Parmentier} \& {Gilmore}(2005)}]{Parmentier2005}
{Parmentier}, G. \& {Gilmore}, G. 2005, \mnras, 363, 326

\bibitem[{{Perets} \& {{\v S}ubr}(2012)}]{Perets2012}
{Perets}, H.~B. \& {{\v S}ubr}, L. 2012, \apj, 751, 133

\bibitem[{{Pfalzner}(2009)}]{Pfalzner2009}
{Pfalzner}, S. 2009, \aap, 498, L37

\bibitem[{{Portegies Zwart} {et~al.}(2010){Portegies Zwart}, {McMillan}, \&
  {Gieles}}]{Zwart2010}
{Portegies Zwart}, S.~F., {McMillan}, S.~L.~W., \& {Gieles}, M. 2010, \araa,
  48, 431

\bibitem[{{Rochau} {et~al.}(2010){Rochau}, {Brandner}, {Stolte}, {Gennaro},
  {Gouliermis}, {Da Rio}, {Dzyurkevich}, \& {Henning}}]{Rochau2010}
{Rochau}, B., {Brandner}, W., {Stolte}, A., {et~al.} 2010, \apjl, 716, L90

\bibitem[{{Traficante} {et~al.}(2015){Traficante}, {Fuller}, {Peretto},
  {Pineda}, \& {Molinari}}]{Traficante2015}
{Traficante}, A., {Fuller}, G.~A., {Peretto}, N., {Pineda}, J.~E., \&
  {Molinari}, S. 2015, \mnras, 451, 3089

\bibitem[{{Vesperini}(1998)}]{Vesperini1998}
{Vesperini}, E. 1998, \mnras, 299, 1019

\bibitem[{{Wall} {et~al.}(2019){Wall}, {McMillan}, {Mac Low}, {Klessen}, \&
  {Portegies Zwart}}]{Wall2019}
{Wall}, J.~E., {McMillan}, S.~L.~W., {Mac Low}, M.-M., {Klessen}, R.~S., \&
  {Portegies Zwart}, S. 2019, arXiv e-prints [\eprint[arXiv]{1901.01132}]

\bibitem[{{Weidner} {et~al.}(2010){Weidner}, {Kroupa}, \&
  {Bonnell}}]{Weidner2010}
{Weidner}, C., {Kroupa}, P., \& {Bonnell}, I.~A.~D. 2010, \mnras, 401, 275

\bibitem[{{Wood} \& {Churchwell}(1989)}]{Wood1989}
{Wood}, D.~O.~S. \& {Churchwell}, E. 1989, \apjs, 69, 831

\bibitem[{{Zonoozi} {et~al.}(2011){Zonoozi}, {K{\"u}pper}, {Baumgardt},
  {Haghi}, {Kroupa}, \& {Hilker}}]{Zonoozi2011}
{Zonoozi}, A.~H., {K{\"u}pper}, A.~H.~W., {Baumgardt}, H., {et~al.} 2011,
  \mnras, 411, 1989

\end{thebibliography}

\end{document}